\documentclass{article}

\usepackage{PRIMEarxiv}

\usepackage[utf8]{inputenc} 
\usepackage[T1]{fontenc}    
\usepackage{hyperref}       
\usepackage{url}            
\usepackage{booktabs}       
\usepackage{amsfonts}       
\usepackage{nicefrac}       
\usepackage{microtype}      
\usepackage{fancyhdr}       
\usepackage{graphicx}       
\usepackage{siunitx}
\usepackage[table]{xcolor}
\graphicspath{{media/}}     
\usepackage{soul}
\usepackage{changes}
\usepackage{lineno}

\title{Emergent self-adaptation in an integrated photonic neural network for backpropagation-free learning
}

\author{
  Alessio Lugnan \\
  Photonics Research Group \\
  Ghent University - imec \\
  Ghent 9052, Belgium\\
  \texttt{alessio.lugnan.1@unitn.it} \\
   \And
  Samarth Aggarwal \\
  Department of Materials \\
  University of Oxford \\
  Parks Road, Oxford OX1 3PH, UK\\
   \And
  Frank Brückerhoff-Plückelmann \\
  Department of Physics, CeNTech \\
  University of Münster \\
  Heisenbergstraße 11, 48149 Münster, Germany\\
  \And
  C. David Wright \\
  Department of Engineering \\
  University of Exeter \\
  Exeter EX4 4QF, UK \\
    \And
  Wolfram H. P. Pernice\\
  Department of Physics, CeNTech \\
  University of Münster \\
  Heisenbergstraße 11, 48149 Münster, Germany\\
    \And
  Harish Bhaskaran \\
  Department of Materials \\
  University of Oxford \\
  Parks Road, Oxford OX1 3PH, UK\\
   \And
   Peter Bienstman \\
 Photonics Research Group\\
  Ghent University - imec \\
  Ghent 9052, Belgium\\
}

\begin{document}
\maketitle
            
\begin{abstract}
Plastic self-adaptation, nonlinear recurrent dynamics and multi-scale memory are desired features in hardware implementations of neural networks, because they enable them to learn, adapt and process information similarly to the way biological brains do. In this work, we experimentally demonstrate these properties occurring in arrays of photonic neurons. Importantly, this is realised autonomously in an emergent fashion, without the need for an external controller setting weights and without explicit feedback of a global reward signal. Using a hierarchy of such arrays coupled to a backpropagation-free training algorithm based on simple logistic regression, we are able to achieve a performance of 98.2\% on the MNIST task, a popular benchmark task looking at classification of written digits. The plastic nodes consist of silicon photonics microring resonators covered by a patch of phase-change material that implements nonvolatile memory. The system is compact, robust, and straightforward to scale up through the use of multiple wavelengths. Moreover, it constitutes a unique platform to test and efficiently implement biologically plausible learning schemes at a high processing speed.
\end{abstract}

\keywords{Neuromorphic computing \and Machine learning \and Self-adapting systems \and Synaptic plasticity \and Silicon photonics \and Phase change materials \and Reservoir computing}

\section{Introduction}
In recent years, computational power and applicability of artificial neural networks (ANNs) have grown rapidly, to the point that this technology is taking a more and more important role in many different fields and aspects of society \cite{lecun2015deep,dong2021survey}. However, the mainstream approach of simulating ANNs in software is highly inefficient because of the large number of parallel operations required for inference and training \cite{strubell2019energy,thompson2020computational,wu2022sustainable}.  On the other hand, biological brains show us that more versatile, more powerful and continuously learning neural networks exist that are extremely energy efficient. Still, there are many unknowns regarding the mechanisms of learning and memorizing in our brain, and today's ANNs models are based on an extremely simplified abstraction of the brain's behaviour \cite{schmidgall2023brain}. 

An example of a research path striving to correct this mismatch is the search for biologically plausible learning rules \cite{schmidgall2023brain,lillicrap2020backpropagation,taherkhani2020review,jeon2023distinctive, hinton2022forward, nakajima2022physical}. This search mainly originates from the evidence that backpropagation (BP), the pillar of conventional training approaches, is not likely to happen in biological neural networks.  Therefore, researchers in the field are looking for biologically plausible learning mechanisms to obtain powerful and efficient ANNs. In particular, plastic self-adaptation is a central property in this regard, as it is considered to be the main enabling mechanism behind memory and learning in biological brains \cite{magee2020synaptic}. We consider a network `plastic' when the response of its components (nodes, connections) to their input depends on the history of this input, in a non-volatile way w.r.t. the relevant timescales. Importantly, thanks to plastic self-adaptation, a suitably designed physical neural network can learn to perform useful functions just by plastically adapting to its inputs in an autonomous way, without the need for an external controller setting the plastic weights. Although there are still no powerful training algorithms nor ANN architectures able to fully achieve this brain-like type of learning on hardware, we agree that plasticity-based learning and self-adaptation is very likely to play a fundamental role in future development of large-scale neuromorphic hardware. In fact, intense research effort is being spent both to achieve it in ANNs \cite{xu2023reconfigurable} and, in parallel, to better understand related mechanisms in biological neural networks \cite{schmidgall2023brain,jeon2023distinctive}. 

An attractive feature of such self-adaptation is that it could alleviate a major scalability issue in hardware implementations of ANNs. Indeed, current state-of-the-art training approaches require full and precise tuning of network parameters and observability of internal states (as it is demanded by BP and gradient descent). In hardware implementations this implies access to the internal components of an ANN through physical connections and control devices, making it extremely hard to actually scale these systems up to numbers of synapses and neurons comparable to the ones in biological brains. However, learning by plastic self-adaptation would remove this limitation, since the network connections would be used at the same time both to process input information and to train the network parameters.

In this work we provide a key step forward in development of autonomous self-adpating neuromorphic computing, by experimentally realising for the first time a scalable hardware ANN whose physical nodes exhibit both volatile memory (short-term plasticity) and non-volatile memory (long-term plasticity), providing at the same time high computing power and the possibility of efficient training through self-adaptation. Importanly, the network's plastic behaviour is fully emergent in the sense that it does not rely on an external controller updating the synaptic weights, or on a global reward signal that is explicitly fed back into the network. Our system is implemented in silicon photonics \cite{lockwood2021silicon}, a compact and industry compatible technology to create chip-based optical networks. In order to realise the plasticity, we use phase change materials (PCMs), whose properties can be modified in a nonvolatile way using optical pulses.

Compared with electronics-based or other neuromorphic computing platforms \cite{markovic2020physics,schuman2022opportunities,christensen20222022}, photonics offers unique advantages in terms of parallelism, energy efficiency, latency and bandwidth of interconnects \cite{shastri2021photonics, pavanello2023special, xu2023reconfigurable}. These are particularly relevant for the development of large-scale hardware ANNs, which comprise a huge number of parallel weighted connections (synapses). Such advantages ultimately arise from the intrinsic difference in the physics behind signal propagation: differently from current-based signals conveyed by an electronic connection, photons travelling through a dielectric medium do not directly interact with each other. This enables the transmission of multiple signals in parallel through the same channel by using light of different wavelengths (i.e. WDM, short for 'wavelength division multiplexing'). This can happen at high speeds and with low energy loss. On the other hand, for the very same reason, nonlinearity and memory have been notoriously difficult to implement efficiently in photonics.

Recently however, phase change materials (PCMs) have been shown to introduce all optical non-volatile memory, and thus physical plasticity, into integrated photonics with relatively high energy efficiency and speed \cite{wuttig2017phase, feldmann2019all}.  In particular, chalcogenide alloys such as GST (short for Ge\(_{\text{2}}\)Sb\(_{\text{2}}\)Te\(_{\text{5}}\)) can be deposited in thin films on top of integrated photonic waveguides, whose optical absorption and refractive index depend significantly on the PCM memory state, which is in turn determined by how much of the PCM is in the amorphous state or in the crystalline state. Specifically, infrared light absorption by crystalline GST is much higher compared to amorphous GST. Importantly, powerful enough optical pulses travelling through the waveguide can quickly heat and melt the PCM layer, whose final non-volatile state will depend on how fast the optical heating decays: slow cooling allows the melted PCM to crystallize, while fast cooling leaves it in the amorphous state. Typical optical pulses used for memory switching have peak powers of a few tens of milliwatts and durations of tens to hundreds of nanoseconds. In this work, we employ GST layers to introduce all-optical cascadable memory, and thus long-term plasticity, in an integrated photonic ANN.

Although photonics and PCMs have been used to build neuromorphic systems before, they mainly rely on an external control scheme that explictly sets the weights. As such, they can be described only as plastic in the very narrow sense that they can be changed, but they lack autonomous emergent behaviour. Moreover, current state-of-the-art approaches still have some additional drawbacks. Indeed, the difficulty of fabricating efficient and cascadable nonlinear nodes is still a major impediment to the scalability of neuromorphic photonics systems \cite{shastri2021photonics}. This challenge has been tackled, for instance, by employing all-optical PCM switching in order to obtain a threshold-like nonlinearity on optical input pulses at different wavelengths \cite{feldmann2019all}. However, this approach requires separate optical pulse generation for the input and output of a neuron, and a dedicated operation cycle to reset the PCM state after a neuron activation, making the employment of many cascaded neurons challenging in practice. In contrast, in this article we present a fully autonomous recurrent neural network capable of processing sequential data, whose nodes concurrently provide nonlinearity, multi-scale volatile memory and plastic self-adaptation.
Another popular approach to build artificial neurons is exploiting the nonlinearity arising from converting optical signals into electric ones by means of a photodetector \cite{tait2019silicon, amin2019ito,nahmias2016integrated}. In order to cascade multiple neurons of this type, the signal can be reconverted to the optical domain by means of a modulator. Nevertheless, this approach presents evident scalability issues, such as a relatively large neuron footprint, high complexity and copious metal wiring. Moreover, similarly to the aforementioned approach based on PCM, every neuron layer requires two dedicated optical input channels. Furthermore, an important general challenge is to cascade multiple neuron layers in a photonic integrated ANN, which can be trained in-situ and online \cite{buckley2023photonic}. Indeed, state-of-the-art efforts towards this direction managed to deploy only a quite limited number of neurons and layers \cite{pai2022experimentally, bandyopadhyay2022single}. Again, as mentioned before, in these artificial photonic neurons, autonomously emerging plasticity is hardly ever considered, especially in the context of scalable networks.
Even outside the field of photonics, e.g. considering the more mature electronics-based neuromorphic hardware, most experimental works about self-adaptive neuromorphic computing are still about single components (such as an artificial neuron or  a synapse) rather than full ANNs \cite{xu2023reconfigurable}. Still, self-adaptation is considered to be an essential challenge and opportunity for future research.

In this work, we present an experimental realisation of plastic photonic neurons in scalable arrays. We combine for the first time the volatile nonlinear dynamics of silicon microring resonators (MRRs) and the non-volatile memory provided by PCM cells, in order to create an autonomously self-adapting dynamical system. In addition, we pair this with a novel cascaded architecture based on simple linear regression, where the most promising plastic adaptations are selected and combined. The training is backpropagation-free and vastly simplified compared to e.g. backpropagation in deep neural networks. Moreover, the system naturally lends itself to WDM exploitation, such that the cascading does not come at the expense of on-chip footprint.  Importantly, the proposed neuromorphic hardware can be trained and used as a testing platform for biologically compatible training procedures based on plastic self-adaptation and recurrent dynamics, as we discuss later on. As a test for its learning and inference capabilities, we show that the system, combined with a novel backpropagation-free training scheme, achieves a particularly good accuracy of 98.2\% on the MNIST \cite{lecun2010MNIST} task, a popular benchmark task looking at classification of written digits.

In Section \ref{sec:PPRRNN} we introduce the proposed type of integrated photonic network and its main properties. In Section \ref{sec:plasticity} we then present an investigation on the emergent network plasticity properties for a highly nonlinear time series classification task, using purposely constructed pulse sequences to trigger plasticity. In Section \ref{fig:MNIST}, building on this knowledge, we subsequently introduce a hierarchical network architecture continuously operating in the plastic regime without requiring specially constructed training sequences. This is coupled to a simple BP-free learning scheme that amplifies the most promising autonomously emerging plastic adaptations. As an example application, we show that the network response can be used to achieve high accuracy of 98.2\% on the popular MNIST benchmark task. In the \textit{Discussion} section, we explore the scalability of the proposed hardware neuromorphic platform and the relation to existent biologically plausible algorithms, like FF (Forward-Forward) and DFA (Direct Feedback Alignment). The Supplementary Document contains further material regarding the single plastic building block (MRR with PCM) and the investigation of the plasticity property.

\section{A scalable photonic recurrent neural network with emergent synaptic plasticity}
\label{sec:PPRRNN}
\begin{figure}[!htbp]
  \includegraphics[width=370pt]{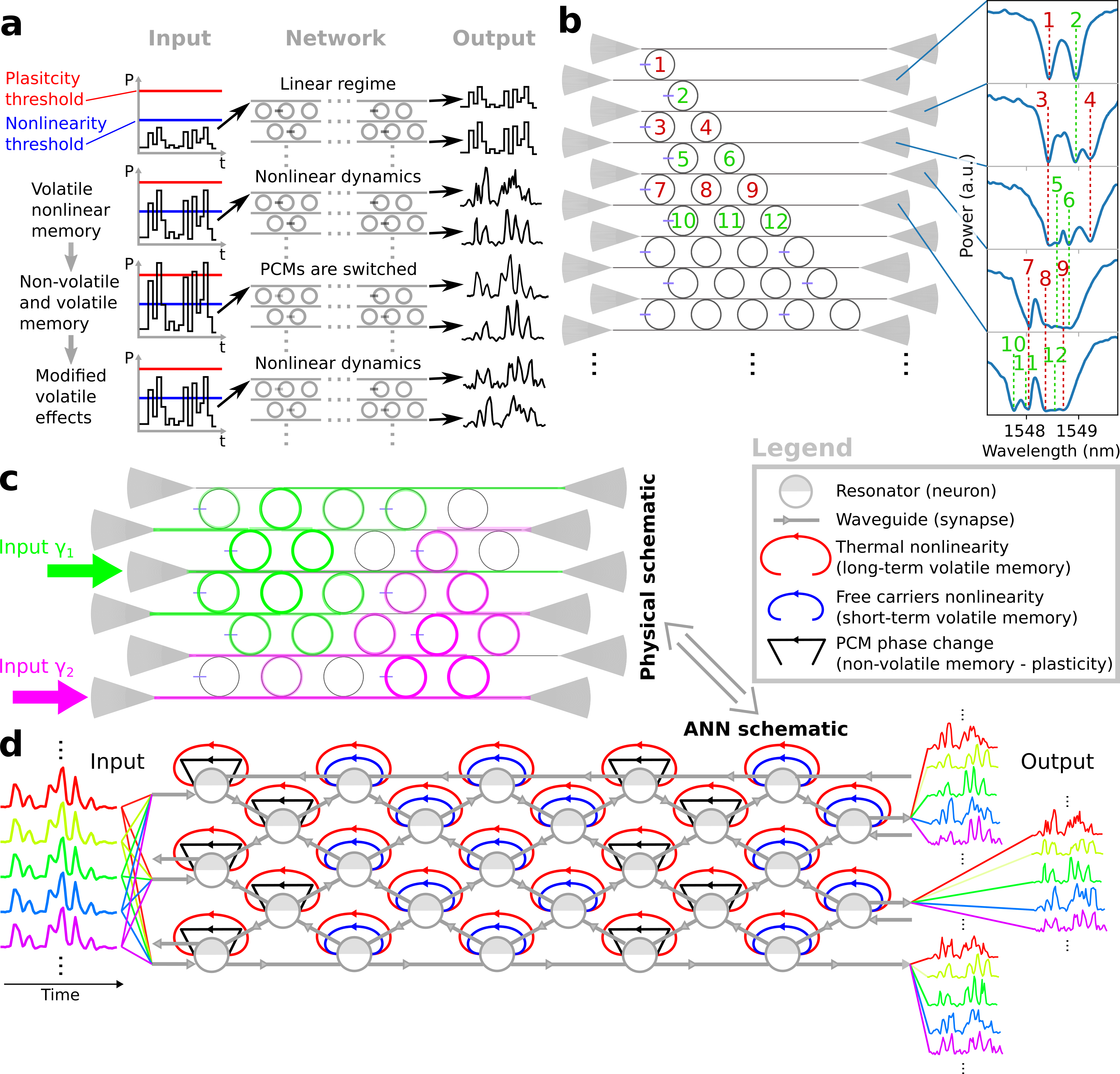}
  \centering
  \caption{\textbf{Photonic plastic recurrent resonator neural network (PPRRNN).} \textbf{a} Basic functionality of a PPRRNN. \textit{First row}: a low-power input waveform does not trigger nonlinear dynamics, thus resulting in a linear network response (no distortion on the output waveforms). \textit{Second row}: a powerful enough input waveform can excite nonlinear dynamics in the PPRRNN, so that different nonlinear representations of the input are obtained at different output ports (and at different wavelengths). \textit{Third row}: a further increase in input power can trigger non-volatile changes in the PPRRNN, due to PCM switching. \textit{Forth row}: setting the same input power as in the second row results in different output nonlinear representations because of the previous non-volatile changes (see third row). These properties enable network training through plastic self-adaptation. \textbf{b} Design of a triangular PPRRNN and measured optical spectra at the indicated output ports, each corresponding to a low-power input inserted through the grating coupler to the left on the same straight waveguide. The numbers indicate which nodes correspond to which resonant dips in the spectra. \textbf{c} Possible light distribution in two `virtual' networks occurring in a rectangular PPRRNN by exciting it with coherent light at two different wavelengths and at two different input ports. \textbf{d} Corresponding equivalent ANN schematic showing optical connections (grey) and recurrent connections associated with different nonlinear effects in the network nodes excited by light propagation (see the legend).}
  \label{fig:NNschematics}
\end{figure}

We present a compact and simple (in terms of design and fabrication) integrated photonic circuit that mimics several key properties of biological neural networks. We now explain its main operational characteristics, arising from a balance between volatile and non-volatile all-optical nonlinear memory. The system takes as input and returns as output multiple time-dependent optical signals. If the input power is high enough (over the \textit{nonlinearity threshold} but below the \textit{plasticity threshold}, see Fig. \ref{fig:NNschematics} a), the corresponding outputs consist of nonlinear transformations with memory (here also referred to as \textit{representations}) of the input, resulting from complex multiphysics dynamics occurring in our photonic network. Such a network activity does not modify the behaviour of the nodes in a persistent way. However, increasing the power above the plasticity threshold, results in nonvolatile changes of the network response, that persist when the power is decreased again below the plasticity threshold. Important to realise is that the exact plastic changes depend on the time evolution of the light intensities inside the different nodes. These in turn depend in a nontrivial way on the input sequence sent into the system, which is subject to all the nonlinear resonances inside the MRRs. As such, we have created a system that can autonomously modify its behaviour in an emergent fashion based on the inputs it receives, and is able to encode this in long-term memory. This way, as we will demonstrate  in Section \ref{sec:plasticity}, multiple and diverse permanent network modifications can be obtained by means of different input signals.

Additionally, our photonic neural network can take in several time-dependent inputs with different optical wavelengths at the same time (WDM), at each physical port, while producing as many output signals at each output port. This greatly increases the network computational power and throughput. Conveniently, in the same circuit and depending on the employed wavelength, different wavelengths can either be coupled together by the photonic neurons, so as to expand the effective network dimension, or they can be processed separately and concurrently by different subnetworks (consisting of disjoint sets of neurons) forming spontaneously, so as to carry out multiple tasks at the same time. In the rest of this section we will go into more technical details in order to explain how the described network properties arise.

The building blocks of the neuromorphic hardware are simple, compact and mature photonic devices, namely silicon microring resonators (MRRs) \cite{bogaerts2012silicon}, which we drive into a nonlinear regime in order to achieve cascadable nonlinear nodes with multi-scale volatile memory \cite{van2012cascadable,mesaritakis2013micro,mancinelli2014chaotic}. To do so, we exploit the competing effect of variations in temperature and free carrier concentration (with timescales of respectivley a few hundred nanoseconds and a few nanoseconds) triggered by optical input signals. At the same time, for the non-volatile memory, we benefit from the power concentration and the enhanced sensitivity to perturbations granted by the resonant behaviour of MRRs. This allows us to achieve a sufficient optical contrast with relatively short PCM patches, thus obtaining more efficient and faster memory operations \cite{lugnan2022rigorous}. That way, we can introduce cascadable and efficient all-optical non-volatile memory nodes into our photonic ANN. In the past, MRRs have been successfully employed to build synapses or neurons in neuromorphic computing applications \cite{biasi2023photonic}, as well as integrated photonic PCM devices\cite{chakraborty2018toward, feldmann2019all, feldmann2021parallel}. In this work, we combine for the first time nonlinear dynamics of silicon MRRs and non-volatile memory provided by PCM cells, to build a hardware ANN with multiscale volatile memory and self-adaptive plasticity.

The network architecture we propose, which we call \textit{photonic plastic recurrent resonator neural network} (PPRRNN), is composed of an arrangement of nonlinear nodes (bare silicon MRRs) and plastic nodes (silicon MRRs with PCM, for details regarding this single component see the Supplementary Document, Section 1), that are coupled to a number of straight waveguides (see e.g. Fig. \ref{fig:NNschematics} a, c).  
Laser light sent into any one of the different straight input waveguides will only couple significantly to those rings along the light path that have a resonance close to the wavelength of the laser. At the same time, these rings will also act as connections to neighbouring straight waveguides, setting up a non-trivial interconnection topology that depends on the wavelength, on the PCM states and, for high enough input power, on the volatile nonlinear effects in silicon. Considering for example a triangular PPRRNN (i.e. one in which the MRRs linking straight waveguides form a triangular arrangement, as in Fig. \ref{fig:NNschematics} b), measurements of spectra in the linear regime, i.e. for low enough input power, show the overlapping of the resonance dips of different MRRs (Fig. \ref{fig:NNschematics} b). Because of light interference, nontrivial spectral features can arise from the coupling of multiple MRRs. 

Although the MRRs are designed to be identical, each one shows a different resonance wavelength due to fabrication imperfections. In this particular chip, resonances are generally red-shifted as the MRR position is moved to the right or upwards, with a smaller random shift superimposed on top of this. Thanks to this correspondence between MRR position in the spatial and in frequency domain, significantly different input wavelengths (i.e. significantly larger than the MRR resonance width) in a PPRRNN can be coupled to different groups of MRRs (see an example in Fig. \ref{fig:NNschematics}c). Operationally, each group of coupled MRRs corresponds to a different virtual network, which can operate separately and in parallel if the corresponding input wavelengths are different enough. Exploiting this property, a PPRRNN can be designed to host a few large networks comprising many coupled MRRs, or many smaller networks that can work separately and in parallel at different wavelengths, even sharing the same input ports. Moreover, many different wavelengths can excite the same group or overlapping groups of nodes, through the different quasi-periodic resonances in a single MRR. These virtual networks are an important ingredient in the scalability properties of a PPRRNN, as they do not require additional chip area.

If the network input has high enough power, the silicon nonlinear effects in the excited MRRs can shift and change the shape of the resonance dips, enabling complex dynamic responses \cite{van2012cascadable, mancinelli2014chaotic}. In particular, temporary resonance perturbations due to free carriers and thermal effects (blue and red shift respectively) provide the nonlinear activation functions of the artificial neurons but also, respectively, short and long term volatile memory. Moreover, those MRRs with a PCM cell (one in every three in each row was chosen for this work) also feature non-volatile memory. In Fig. \ref{fig:NNschematics}d we depict the ANN diagram corresponding to the physical PPRRNN in Fig. \ref{fig:NNschematics}c, showing the main connectivity and memory elements, leaving out the dependence of the neuron response to the input wavelength. In this work we consider the optical connections (grey arrows) as instantaneous w.r.t. the dynamics of input signals and memory effects (recurrent arrows), since light propagation happens much faster. Therefore, the memory effects of the nodes are in practice applied to the equilibrium state of the purely optical network dynamics. 

It should be stressed that the plastic nodes (MRRs with PCM) have less pronounced nonlinear and volatile memory effects than bare MRRs, due to the lower Q factor caused by the optical loss at the PCM cell. Therefore, in Fig. \ref{fig:NNschematics}d we neglect the weaker memory due to free carriers (blue arrows), while temperature still has a significant but reduced influence.
In this section we have introduced a triangular PPRRNN in order to show how the spectrum of an increasing number of coupled nodes builds up. However, from now on, we will only consider rectangular PPRRNNs, which are more compact.

\section{Emergent synaptic plasticity enables self-adaptive non-volatile weight modifications without external control} \label{sec:plasticity}
In this section, we present an experimental investigation on self-adaptation due to the emergent plasticity property in a PPRRNN, and on how it can be exploited to improve machine learning (ML) performance on a time-series classification task, without explicitly tuning the network weights externally. Here, our main aim is to demonstrate that non-volatile plasticity in our network is \textit{rich}, \textit{accessible} and can be \textit{concurrent} with volatile nonlinear memory. By \textit{rich}, we mean that multiple and significantly different non-volatile plastic configurations can be realised by slightly different input optical waveforms. By \textit{accessible}, we mean that these plastic configurations can be obtained using reasonable time-dependent optical input signals (not too powerful, not too noise-sensitive response, not too slow or fast, etc...) and affect the network output in a well-readable way. \textit{Concurrency} with volatile nonlinear effects means that the network is able to exploit both non-volatile (plastic) and volatile memory at the same time, in order to carry out a task. Indeed, richness, accessibility and concurrency with volatile effects are three fundamental properties for physical plasticity in order to be practically employable for biologically plausible learning based on self-adaptation. Here we demonstrate, for the first time to the best of our knowledge, all three attributes to be readily available in a photonic hardware.

\begin{figure}[!htbp]
  \includegraphics[width=470pt]{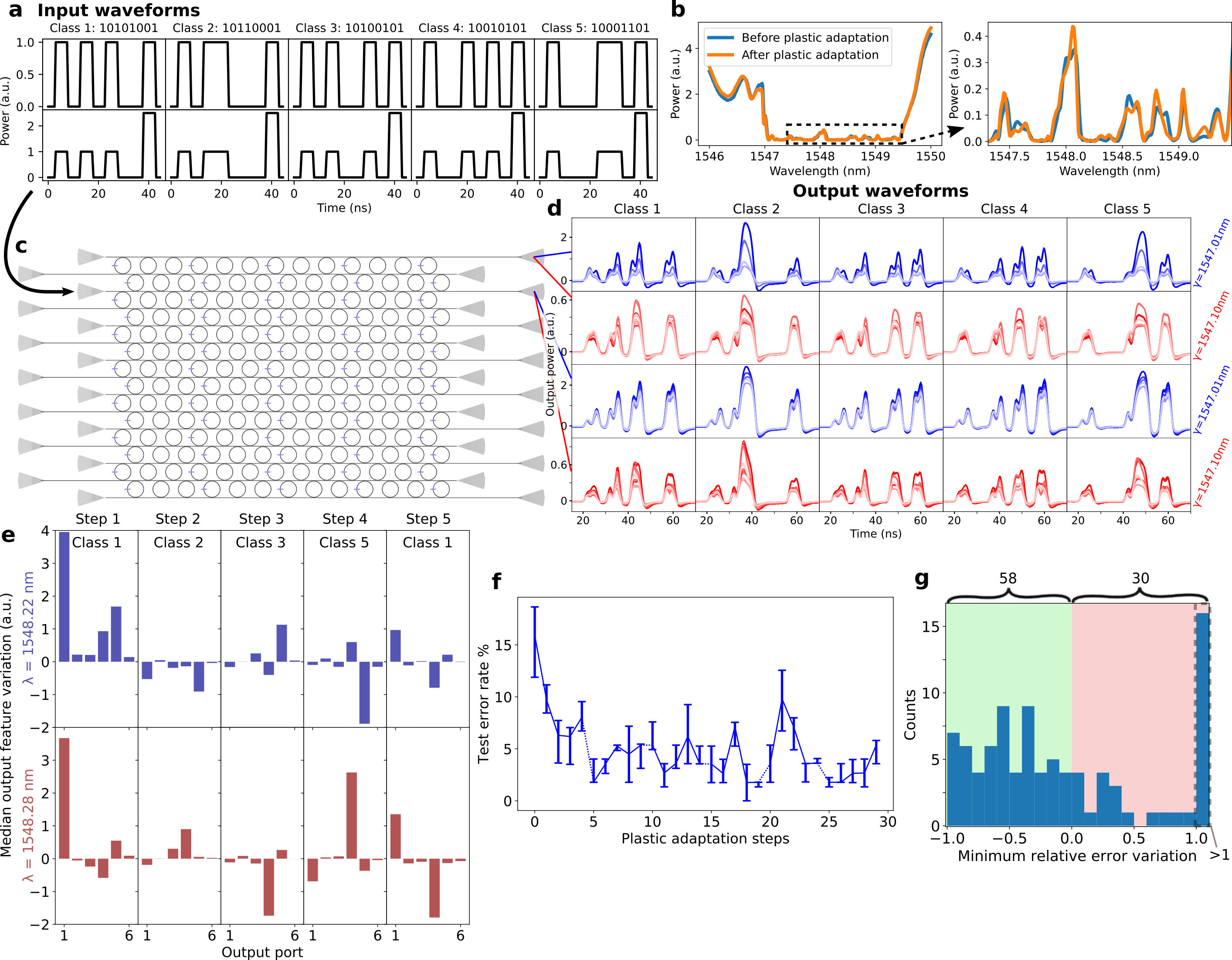}
  \centering
  \caption{\textbf{Impact of plasticity on the network response.} \textbf{a} Example of input waveforms from the considered 5 classes (columns), employed in the inference step (first row) and in the plastic adaptation (PA) step (second row). \textbf{b} Example of non-volatile spectrum modification due to the plastic adaptation of the considered PPRRNN (measured using the third input port and the third output port). \textbf{c} Schematic of the considered PPRRNN. \textbf{d} Examples of average output waveforms (median, 10\% and 90\% percentiles) for different input waveform classes (columns) at two different ports and two different wavelengths (rows). The five different shades in the plots correspond, from dark to light, to the output obtained at the beginning and after different subsequent PA steps from class 1 to 5. \textbf{e} Example of variations of the median output features (last pulse energy of output waveforms) for different output ports, due to different classes of consecutive PA steps (columns) and for two input wavelengths (rows). \textbf{f} Example of error rates (for the considered 5-classes waveform classification) in a PA step sequence, as a function of consecutive PA steps. \textbf{g} Histogram of the minimum error rate relative variations w.r.t. the initial error, in each measured PA step sequence. This provides an idea of how often and how strongly the PA step sequences improves (negative values on the x axis) the corresponding initial ML performance.}
  \label{fig:plasticity}
\end{figure}

In this section we study an example application in order to gain insight into the plasticity properties of our network, by employing a bespoke task and a dedicated optical training scheme, which allows us to study the system in a more controlled environment. For the task, we consider 5 classes of input waveforms, each being a different permutation of 4 high bits over 8 bit positions in time (see upper plots in Fig. \ref{fig:plasticity}a). These waveforms have the additional constraint that, at the first and the last positions, bits are always high. Altogether, this represents a temporal frame. A single bit is $\SI{5}{\nano\second}$ long. During the inference phase, a high bit has a relatively low peak power of around $\SI{7}{\milli\watt}$. The pulse power is chosen so that it can trigger nonlinear volatile effects (thus the output waveforms present nonlinear distortions w.r.t. the input) but not significant non-volatile changes. That is, dimensionality expansion can be obtained while the solid-state phase change of the GST patches is considered negligible. Under these conditions, we can test the ML performance for a certain fixed configuration of the non-volatile weights in the network. Fig. \ref{fig:plasticity}d shows examples of average output waveforms corresponding to the five classes (columns), for two different wavelengths (blue for 1549.01 nm and red for 1547.10 nm) and at output ports in rows 1 and 3, with reference to the PPRRNN in Fig. \ref{fig:plasticity}c. Here, it should be stressed that, in order to give an idea of the noise and of potential instability in the acquired network output, we plotted for each output waveform and in the same colour the median and both the 10\% and 90\% percentiles, from a sequence of 5 repetitions of a certain input. The different shades of blue and red, instead, show to the output after different plastic adaptations of the network, which will be discussed later on in this section.

We employed a simple ML pipeline to evaluate the network performance for different plastic configurations: for each input waveform, we applied a linear classifier (logistic regression) on only a single value per output waveform. This value is the output energy corresponding to the last pulse of the input waveform at the considered output port. This particular choice for a small number of features makes the ML task far more difficult. Importantly, without the volatile memory and the nonlinearity provided by our PPRRNN, it is in principle impossible for the employed classifier to learn the classification, since the last transmitted pulse would be independent on the previous pulses. (For clarity, here we stress that a single classifier is applied to all the network outputs, as opposed to the ML model described in the next section, where multiple linear classifiers are applied each to a single output waveform and then combined). 

In order to investigate the variations of the network response and of the corresponding ML performance due to plastic PCM weights modifications, we alternate so-called \textit{inference} and \textit{plastic adaptation steps} (from now on we will shorten the latter to \textit{PA steps}), which we now explain. 

During an inference step, the waveform classes are repeatedly inserted into the PPRRNN, one after another but always separated by around $\SI{2}{\micro\second}$ to eliminate thermal memory between waveforms. The energy of the last pulse of the obtained output waveform is used as the sole feature for the linear classifier. 

During a PA step, a modified version (called here \textit{pump waveform}) of one waveform class is repeatedly inserted. In particular, a pump waveform from a given class is the same as a normal input waveform from the same class, with the only difference that its last pulse has significantly larger peak power (usually by a factor 2.5, see lower plots in Fig. \ref{fig:plasticity}a). The enhanced power in the last pulse is chosen so that it can significantly modify the accessed plastic PCM weights in the network. On the other hand, the first three pulses of a pump waveform are meant to set the same volatile PPRRNN configuration obtained by the first three pulses in a corresponding normal waveform, so that the last enhanced pulse reaches the same nodes and output ports as the last pulse in a normal waveform. This way we aim to demonstrate \textit{richness} of plasticity, showing that different classes of pump waveforms can modify the plastic weight configuration in different ways, related to the specific light path induced by the corresponding class of normal input waveforms. It should be specified that both inference and PA steps consist of a repetition of the waveform  with a total duration of around $\SI{1}{\second}$. Therefore, given the large number of inserted waveform copies (hundreds of thousands), the plastic weight configuration is considered to have reached an equilibrium non-volatile state after each PA step, depending on the class used in the step and also on the order of the classes used in previous PA steps. Fig. \ref{fig:plasticity}b shows an example of non-volatile spectrum modification due to plastic adaptation of the considered PPRRNN. Further practical details regarding the plasticity investigation and the ordering of classes during training can be found in the \textit{Methods}, Section \ref{subsec:methods_plasticityInvestigation}.

A first result we will demonstrate in this section is that different input (pump) waveforms can achieve significantly different plastic weight configurations. Therefore, we now analyse the output of our physical network, without considering any specific ML task. An example is given by Fig. \ref{fig:plasticity}d, where the five different shades in the plots correspond, from darkest to lightest, to the average output waveforms obtained at the beginning and after different subsequent PA steps, spanning over the different waveform classes. We can notice that the output variations due to the rearrangement of internal plastic weights by PA steps are substantial and easy to distinguish. Thus, this shows that network plasticity is well \textit{accessible} in our PPRRNN. Moreover, all the output waveforms present evident and different nonlinear distortions w.r.t. to the input shown in the first row of Fig. \ref{fig:plasticity}a. Since each curve corresponds to measurements for a fixed nonvolatile configuration of the plastic weights, these distortions are due to volatile nonlinear effects. This demonstrates that we can achieve non-volatile plastic adaptation \textit{concurrently} together with volatile nonlinearity.

The output variation due to plastic adaptations (represented, as we have just discussed, by the different shades in the plots of Fig. \ref{fig:plasticity}d) is more accurately quantified in Fig. \ref{fig:plasticity}e, where the bar plots show the median variation of each output feature (i.e. the intensity of the last pulse) for the different output ports (on the x axis). Each plot column shows the variation due to a different PA step, in chronological order, for two wavelengths (rows). It can be easily noticed that different variations are obtained after each different PA step, implying that different plastic weight configurations are achieved in the PPRRNN. Thus, this demonstrates \textit{richness} of network plasticity in our PPRRNN. A more detailed discussion of richness using more extensive experimental results is given in the Supplementary Document, Section 2.

In the previous paragraphs we have discussed plastic effects by directly looking at the network output. Now we take a step further and see how network plasticity can be useful in ML.
In particular, we repeatedly evaluate the aforementioned waveform classification task after different PA steps. Mainly, we are interested in seeing how different plastic weights configurations result in different ML performance, i.e. in different performance of the PPRRNN when employed to provide useful data representation to be fed to a linear classifier. Indeed, we will see that sequences of PA steps allow us to explore the plastic weights' configuration space so that it is likely to achieve significant ML improvements w.r.t. the initial (unadapted) state.
An example of how the repetition of different PA steps can greatly improve the ML performance (regarding the classification of the five different bit patterns) is shown in Fig. \ref{fig:plasticity}f, where the initial error rate in a PA step sequence is decreased by more than a factor 6. Remarkably, after the improvement due to the first six PA steps, the error rate stays significantly lower than its initial value for the subsequent steps. This result shows that our approach to modify the plastic weight configuration of the PPRRNN can permanently improve the linear separability of the output feature values. Importantly, this is achieved without any externally supervised weight training, but just by letting the plastic network adapt to its input, in an emergent fashion. In this way, the proposed method resembles to the way biological brains memorize and learn. 

However, not every measured PA step sequence (each distinguished by a different wavelength or input port, see Table \ref{tab:meas_sess}) resulted in such an evident and stable performance improvement. Nevertheless, we will show that they still allow one to explore the plastic weight configuration space such that ML performances are often significantly improved. In order to illustrate this, we look at how much the ML error decreases w.r.t. the initial network state, as the consequence of subsequent PA steps. For example, considering the PA step sequence evaluated in Fig. \ref{fig:plasticity}f, we are interested in the improvement corresponding to the minimum error achieved w.r.t. the initial error value. In practice, for each measured PA step sequence, we calculate the minimum error rate variation (which is negative if there is a classification improvement) relative to the corresponding initial error rate value, where the minimum is taken over the error values achieved by all the PA steps in the sequence. We calculated this minimum (relative) error variation for each measured PA step sequence, and plotted them in a histogram (Fig. \ref{fig:plasticity}g). It can be noticed that most of the PA step sequences allow to improve the classification performance (shown by negative values on the x axis). Importantly, the distribution does not decrease as the values on the x axis approaches -1 (which corresponds to a complete removal of the initial error thanks to the PA step sequence). This shows that strong performance improvements, as a result of a PA step sequence, are roughly as frequent as small ones. 

It should be stressed that, to demonstrate a full training procedure that optimizes the plastic weights in the PPRRNN independently of the initial conditions of the network (e.g., given by the arrangement of the nodes' resonant wavelengths w.r.t. the input wavelength), is considered as an ambitious goal for a future work. In the next section, instead, we show that different output waveforms (that are nonlinear representations of the input) can be exploited as-they-are by a suitable ML procedure, achieving high accuracy in a far more complex benchmark task (handwritten digits classification).

\section{Combining parallel temporal representations for improved machine learning performance (MNIST classification)}
\label{sec:MNIST}
We will now proceed to present a more universal and relevant benchmark task, namely the 10-class image classification problem from the MNIST dataset for handwritten digits \cite{lecun2010MNIST}. Here we employ a more practical learning scheme that needs no modification of the input waveform to enable plasticity. Additionally, to increase robustness and to better exploit the computational power and multiplexing capabilities of our hardware, we introduce a hierarchical scheme consisting of many networks, where each network is trained to improve upon the performance of the previous one. Importantly, the elements in the hierarchy do not need to be separate structures, but can be different virtual structures in the same network, realised by changing ports and wavelengths.

\begin{figure}[!htbp]
  \includegraphics[width=470pt]{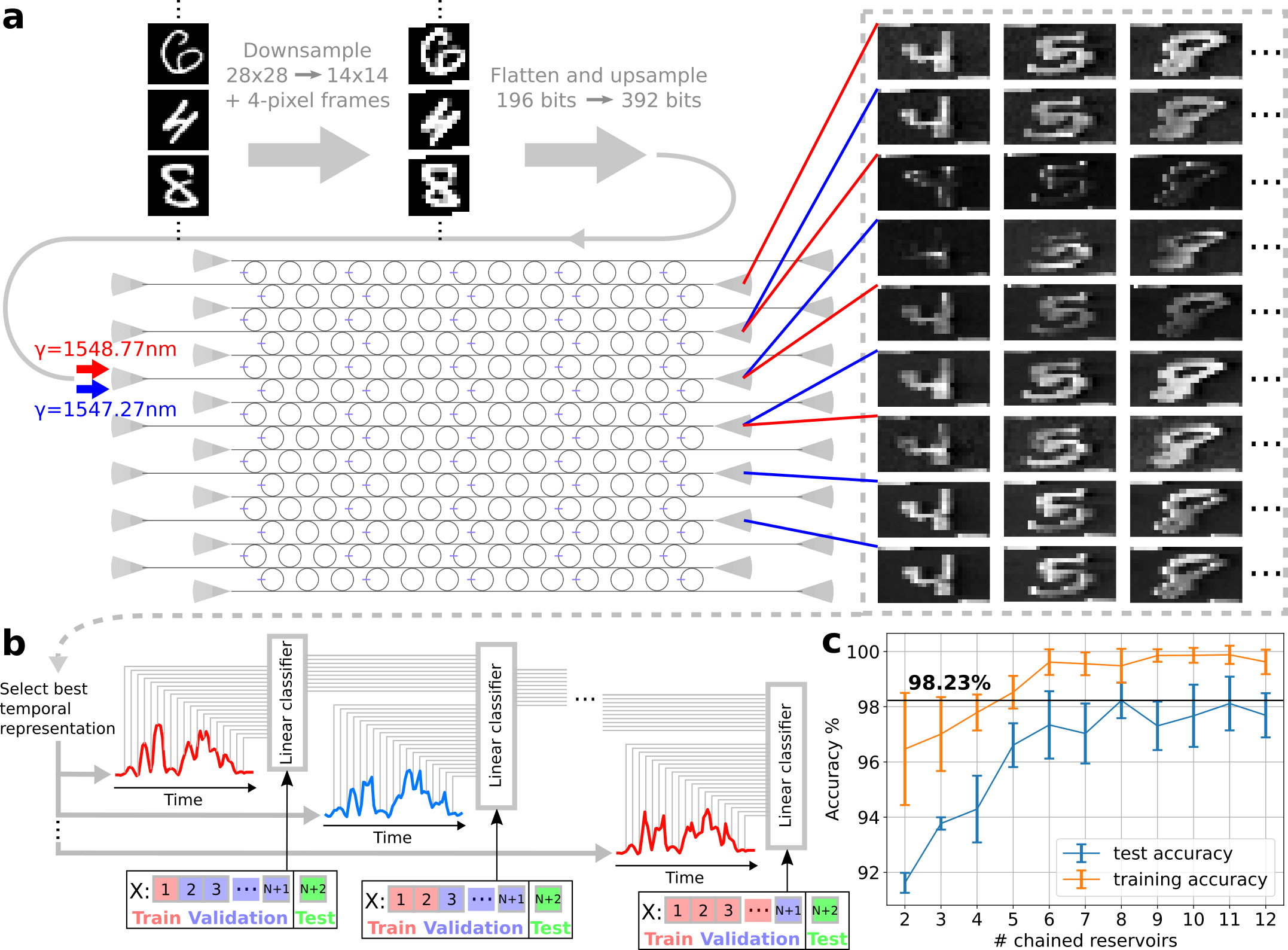}
  \centering
  \caption{\textbf{MNIST images classification combining different output representations from a PPRRNN.} \textbf{a} Example of images insertion into a PPRRNN and of the resulting output representations (in this figure these are spatially rearranged to enable visual intuition, but in truth each is a 1D time-dependent output). After downsampling, flattened images are inserted into a single port (grating coupler on the left) using two wavelengths. Nine different nonlinear representations of the flattened images are then obtained from different output ports and input wavelengths. \textbf{b} Linear ML classifiers are applied to the output representations and combined using a \textit{chaining} ensemble method. \textbf{c} Classification accuracy as a function of the number of chained representations and classifiers. The best average test accuracy obtained is 98.2\%.}
  \label{fig:MNIST}
\end{figure}

We start by flattening each image in the MNIST dataset and insert it as a single time-dependent input that can trigger volatile memory and non-volatile self-adaptation concurrently in a rectangular PPRRNN. Very little preprocessing was employed, motivated exclusively by limitations of the experimental setup (see Methods Section \ref{sec:methods_MLaspects} for more details). As discussed in the previous section, different nonlinear output representations of the input waveform can be obtained in parallel from different physical output ports and for different input wavelengths. Each representation is a waveform (a time-dependent 1D signal) that can be reassembled into a corresponding flattened image of \(n\) pixels (see examples of MNIST images output representations in Fig. \ref{fig:MNIST}a). Differently from the application of kernels in a convolutional neural network (CNN), the generation of these representations is not the outcome of an external learning algorithm, but it depends on the emergent plastic and volatile properties of the PPRRNN and on how the input is inserted (power, wavelength, bitrate). Here we will show that an ensemble of linear classifiers (multi-class logistic regressors) can learn how to synthesize and combine the information unveiled in these representations, in order to greatly improve classification performance, while still having a backpropagation-free lightweight machine learning (ML) pipeline.

In particular, we considered each parallel PPRRNN's representation of the input samples as the output of a stand-alone \textit{reservoir}, according to the \textit{reservoir computing} (RC) framework\cite{tanaka2019recent,lugnan2020photonic}. RC is a hardware-friendly ML approach where only a linear model (a single-layer ANN) is trained, and is applied to the output of a fixed nonlinear dynamical system (the reservoir, e.g. a recurrent neural network with fixed synaptic weights), that is in turn excited by a time-dependent input. Following this scheme, we applied and trained a linear classifier (in software) on each PPRRNN's representation, forming an RC system per representation. Each single RC system alone, though, is a rather weak classifier (the accuracy does not exceed 88\%, which is the accuracy of a linear classifier directly applied on the input, without any neuromorphic hardware). However, we could assemble a much stronger classifier by combining the RC systems together in a special way (see Fig. \ref{fig:MNIST} b and Section \ref{sec:methods_MLaspects} in Methods for technical details), such that each RC system is be trained to correct the errors made by the previous reservoir. This  scheme exploits the PPRRNN's capability of efficiently producing several different nonlinear representation of a temporal input at the same time, leaving to the electronics only the linear models, which are easy to train and computationally cheap. Importantly, the training procedure does not require gradients, is backpropagation-free and provides an example of how the PPRRNN can be employed for powerful biologically-plausible neuromorphic computing.

Fig. \ref{fig:MNIST}c shows the obtained test and training accuracy averaged over the cross-validation loop, as a function of the number \(N\) of classifiers chained together. The error bars represent the standard deviation of the cross-validation accuracy. It can be noticed that our chaining method significantly improves the classification accuracy and reduces overfitting as the number of chained representations increases, until the improvement saturates. We obtained a maximum average accuracy of 98.2\(\%\) from a chain of length 8. Importantly, a much lower maximum accuracy (around 92\%) was reached if we employ a more straightforward method of combining the representations, namely stacking all the features together to obtain a large spatio-temporal representation on which the readout was trained. This is probably caused by overfitting because of the larger number of features in a single training. On the other hand, we believe significant improvement in maximum accuracy for both methods could still be achieved just by measuring more samples to employ for training.

Finally, we compare our best average classification accuracy with the ones experimentally demonstrated in other works about photonic neuromorphic hardware (see Table \ref{tab:table_MNIST}). We achieved a high accuracy compared to other works, demonstrating a high computational power and stability of the plastic spatio-temporal representations produced by our photonic network. Moreover, the two works outperforming our accuracy level rely heavily on powerful feature extraction performed in software. 

In fairness, the aim of the compared works is often to improve also on performance parameters other than accuracy, like energy efficiency or throughput per footprint area. However, these other parameters are usually expressed in terms of number of multiply-accumulate operations (MACs) that can be trained and performed by an ANN. Since we employ a hardware-based dynamical system approach, it makes little sense to directly compare e.g. energy efficiency in terms of MACs/J, because the operations in our recurrent neural network are not externally programmable, although they can be reconfigured via plastic self-adaptation. Nevertheless, in order to give a quantitative idea of the computational throughput per chip area and of the energy efficiency of a PPRRNN, we approximately estimated \(10^{15}\) (MACs+NLOs)/s/mm\(^2\) and \(5\times10^{15}\) (MACs+NLOs)/J respectively (see Section \ref{sec:methods_energyConsumption} under Methods for details on this estimation). The units for these quantities are similar to the usual energy and aerial efficiency estimations (MACs/s/mm\(^2\) and MACs/J respectively), but imply that each MAC operation is also accompanied by a nonlinear operation (NLO), since we work with nonlinear neurons. It should be stressed that the nonlinearity in each node in principle enhances the computational power of our system w.r.t. to photonic linear accelerators, although our synaptic weights cannot be precisely and individually programmed. Moreover, in our estimation we neglected the operations happening in the hidden recurrent layers in our PPRRNN, so as to make it somehow more comparable with linear accelerators, which perform one single layer of weighted connections at a time. Even so, the estimated computational throughputs and efficiencies for our PPRRNN are well beyond those of photonic neuromorphic hardware where synaptic weights are precisely programmable \cite{shastri2021photonics,dabos2022neuromorphic}. Furthermore, it should be noticed that it is straightforward to significantly increase aerial and energy efficiencies by using MRRs with lower radii, higher Q factors and shorter GST patches. Further details regarding energy consumption, footprint and throughput of the employed PPRRNNs are discussed in \textit{Methods}, Section \ref{sec:methods_energyConsumption}.

\rowcolors{2}{gray!15}{white}
\begin{table}
\caption{Comparison of MNIST classification accuracy experimentally demonstrated employing photonic neuromorphic hardware.}
  \centering
  \begin{tabular}{p{3.5cm}p{5.5cm}p{4.5cm}p{1.5cm}}
    \toprule
    Work & System & ML approach & MNIST accuracy \\
    \midrule
    This work & Integrated recurrent ANN based on photonic resonators, with combination of linear classifiers applied on each time-dependent output. With simple preprocessing (mainly downsampling). & Ensemble (chaining) of RC systems based on logistic regression. & 98.2\% \\
    Nakajima, Mitsumasa, et al. (2022)\cite{nakajima2022physical}. & FPGA-assisted fiber-optic system implementing optoelectronic time delay RC. & Deep RC, trained with augmented direct feedback alignment (DFA). & 97.80\% \\ 
Mourgias-Alexandris, G., et al. (2022)\cite{mourgias2022noise}. & CNN (in software) fed into two final photonic layers. & Standard BP on software using a noise-aware training model. & 99.3\% \\ 
Zhou, Tiankuang, et al. (2021). & 3-layer ANN employing large-scale optoelectronic diffractive processing units. & Training in software using BP and adaptive training steps of the optics to adjust for the experimental error. & 96.6\% \\
Feldmann, Johannes, et al. (2021)\cite{feldmann2021parallel} & On-chip photonic crossbar array with PCM performing matrix-vector multiplication for CNN acceleration and a fully connected layer in software. & Standard BP in software. & 95.3\% \\ 
Antonik, Piotr, et al. (2019)\cite{antonik2019large} & Laser, SLM and camera providing a large-scale optoelectronic ANN layer. & RC on extracted features (histograms of oriented gradients, in software) & 98.97\% \\
 Nakajima, Mitsumasa, el al. (2021)\cite{nakajima2021scalable} & On-chip recurrent, passive and coherent photonic network. & RC (spatiotemporal). & 91.3\%  \\
 Bai, Bowen, et al. (2023)\cite{bai2023microcomb} & CNN, where the linear part of the convolutional layer is performed by on-chip photonic devices and circuitry. Two fully connected layers follow in software. & Standard BP and in-situ calibration based on gradient descent control. & 96.6\% \\
  Zhu, H. H., et al. (2022)\cite{zhu2022space} & 2-layer ANN, each layer based on a programmable integrated photonic network, with preprocessing. & Standard BP on software. & 91.4\% \\
    \bottomrule
  \end{tabular}
  \label{tab:table_MNIST}
\end{table}

\section{Discussion}
\subsection{Mechanisms and properties for high scalability}
In Section \ref{sec:plasticity} we have shown that different plastic weight configurations could be achieved in a PPRRNN by insertion of different time-dependent optical signals. Moreover, these non-volatile modifications could often improve the system performance on a simple time-series classification task. Since plastic adaptation is a key mechanism for learning and memory in biological brains \cite{magee2020synaptic}, learning with physical plasticity is an extremely relevant research direction for the development of neuromorphic computing. Indeed, developing a physical platform where dynamics, nonlinearity, volatile and non-volatile memory coexist in complex scalable networks is important in order to provide a physical and experimental underpinning to such a research effort, which is nowadays mainly limited to simulations or entirely externally implemented learning rules \cite{schmidgall2023brain}. In fact, simulating large-scale dynamical systems is arduous and requires large computational resources and simplistic approximations. For instance, approximating or neglecting constraints or richness of response found in physical systems, might prevent the discovery of important learning mechanisms in biological neural networks.

In this section we present the main aspects and properties enabling scalability (in terms of network computational power and size) of the proposed neuromorphic computing approach. Here we would like to highlight the differences between our approach, without externally tunable parameters and based on linear classifiers like in reservoir computing (RC), and the more common one, based on employing backpropagation (BP) on a simulated version of the ANN, whose trained parameters can be transferred to the hardware network via external tuning and correction of hardware non-idealities. In our approach we have reduced control and configurability, and in particular we give up the possibility to accurately predict (before fabrication) and simulate the response of our network, which would have allowed us to estimate the gradient of the cost function and exploit powerful training methods based on BP. In turn, we gain in complexity (having a relatively large number of highly dynamical and nonlinear nodes with multi-scale memory), in robustness to fabrication errors and in low footprint. Moreover, by introducing the all-optical plasticity given by PCM cells in our network, we aim to mitigate the loss in control by potentially allowing for a more biologically plausible way to optimize network parameters, through plasticity and emergent self-adaptation.

\paragraph{Synaptic weight modification without external connections.}
As mentioned before, today state-of-the-art ANNs are trained using BP, which is considered not biologically plausible \cite{taherkhani2020review, lillicrap2020backpropagation,schmidgall2023brain}. BP requires full observability of the neuron states and full tunability of parameters, such as synaptic weights. In practice, where neuromorphic hardware is concerned, this usually requires physical connections in order to observe states and to update weights, so as to apply a training algorithm that runs on an external computer. However, this obviously undermines the scalability of physical ANNs, preventing the use of a large number of neurons and synapses. In this work, instead, plastic weights are modified in a more biologically plausible way through self-adaptation, by exciting the input ports of the network, without requiring dedicated connections. An additional potential advantage of this approach is that, since the signal modifying the plastic weights is inserted at the network input and travels through the normal network connections, the updating of the plastic weights naturally contains information regarding the state of previous nodes and links along the activation path.

\paragraph{Dynamics-enabled cascadability.}
Cascadability of nodes, and also of plastic connections in our case, is critical for scalability of ANNs. However, it is often difficult to achieve in hardware implementations (especially in photonics) without employing a large number of amplification stages or alternative signal sources to compensate for propagation losses, and this may strongly limit scalability. This problem is mitigated when suitably using silicon MRRs as a dynamic node. Indeed, let us consider a PPRRNN (see Fig. \ref{fig:plasticity}c) with rows containing many MRRs in series, assuming for simplicity that the resonant wavelengths are aligned. A non-resonant optical input pulse will reach the corresponding direct output port with potentially negligible energy loss. Instead, a resonant input pulse will be totally or partially absorbed by the first encountered node. However, if the nodes are suitably designed and if the pulse carries enough power, the resonance wavelength of the first node will be red-shifted out of the way by heating due to optical absorption. Therefore, a subsequent pulse can reach the second MRR, whose resonance can be shifted as well, and so on. This way, a detectable signal can reach an output port even when the transmission in the linear regime through the original optical path is very low. 

\paragraph{Parallelism and network expansion via wavelength division multiplexing.}
As discussed in Section \ref{sec:PPRRNN}, even if each node has only four physical connections, these can convey several different signals in parallel and independently by means of WDM, thus greatly expanding the number of network connections, as well as the number of input and output ports. Thanks to the quasi-periodic resonances in the spectrum of a MRR, several different signals at different wavelengths can nonlinearly interact through the silicon nonlinear effects in the optical cavity. Indeed, a powerful enough resonant pulse will simultaneously modify (either temporally or permanently) all the resonances in the MRR spectrum, thus changing the way pulses at other wavelengths excite (or are transmitted by) the neuron. In practice, this mechanism expands the fan-in and fan-out properties of both the artificial neurons and the plastic nodes, whose activation can be achieved by the total power carried by pulses at different times, at different wavelengths and at different physical connections (in the latter case, optical interference comes into play as well).

\paragraph{Multi-timescale computation.}
Nowadays, a major challenge in the development of neuromorphic computing platforms for edge computing is the need to match the timescales of the computing system with the ones of the input information, which may depend on e.g. the type of physical quantities targeted in smart sensing applications \cite{jaeger2021dimensions}. The PPRRNN  proposed here presents dynamic responses with multiple timescales, which can potentially be expanded or controlled. In particular, the fastest timescale is given by the travel time of light signals through the network, considering also that MRRs accumulate resonant light with typical transient times of tens of picoseconds. This timescale can be controlled and extended by choosing the Q factors of the MMRs, or, more effectively, by introducing optical delay lines in the photonic circuit \cite{vandoorne2014experimental}. The second fastest timescale is given by silicon nonlinear effects related to free carrier concentration in ring waveguides, of which time constants can range from a few to tens of nanoseconds. These can also be controlled by applying a suitable p-n junction to the ring waveguide \cite{you200812}. The slowest timescale is provided by the thermo-optic effect either in ring waveguides or in the PCM layer, with characteristic times of hundreds of nanoseconds. Next, the thermal timescale is mainly governed by the temperature dissipation, which can potentially be tuned by design of the photonic circuitry. Interestingly, the combination of effects due to free carriers and temperature can generate self-sustained dynamics in MRRs (self-pulsing) capable of complex and chaotic behaviour \cite{mancinelli2014chaotic}. By suitably tuning the excitation parameters, the decay time of this type of network response can in principle be controlled and extended to slower timescales. Finally, the non-volatile all-optical memory introduced by PCM, allows our PPRRNNs to couple signals inserted at arbitrarily distant times enabling, for suitable types of input excitation, timescale-invariant computation.

\paragraph{High-throughput generation and large choice of data representations within a low footprint.}
As discussed in Section \ref{sec:MNIST}, a large number of different nonlinear representations of an input time series can be achieved by a PPRRNN with a relatively low footprint (e.g. $\SI{0.5}{\milli\meter}^2$), by considering different input and output ports, wavelengths, plastic weights configurations and even input signal parameters (such as bitrate and power). Therefore, once the circuit is fabricated, it is possible to explore its response so to find the parameters providing representations suitable for the considered application, e.g. employing an RC-based approach as we did in Section \ref{sec:MNIST}. Moreover, several representations can be obtained in parallel at different output ports. Furthermore, the number of parallel representations can be multiplied by considering different input ports and different enough wavelengths (see Section \ref{sec:MNIST}). This possibility in principle allows for the generation of hundreds of different representations in parallel within footprints of the order of $\SI{1}{\milli\meter}^2$, enabling high-throughput neuromorphic computing.

\subsection{Relation to biologically plausible in-situ training methods}

Finally, let us briefly discuss the links between our photonic neuromorphic system and two inspiring training approaches aiming at biological plausibility and simplicity of implementation in hardware. In \cite{hinton2022forward}, a surprisingly powerful learning procedure (called \textit{Forward-Forward} algorithm, or FF algorithm) is presented, which replaces the forward and backward passes of BP-based training by two forward passes, with the only difference being the inserted data. This makes the algorithm more biologically plausible and eliminates the BP requirement of accurately knowing all the operations performed by the network. Therefore, the FF algorithm can be implemented in hardware implementations of neural networks, where internal operations are mostly unknown due to the variability arising from fabrication errors and due to complex nonlinear responses of the nodes. Similarly, in this work (Section \ref{sec:plasticity}), there is no backward pass and we modify the network parameters directly by inserting specific input signals, leveraging intrinsic physical plasticity rather than an external learning rule. Another similarity is that in both cases several linear classifiers are trained and then combined (see Section \ref{sec:MNIST}), although in a different way. In the cited article, it is also stressed that \textit{mortal computation}, i.e. computation learned by non-reproducible hardware like the one here presented, may generally allow for higher energy efficiency and lower fabrication costs.

A second relevant work is \cite{nakajima2022physical}, where a hardware-friendly augmented version of \textit{direct feedback alignment} (DFA, see \cite{nokland2016direct}) is presented. DFA already takes a big step towards biological plausibility and on-hardware implementability, by removing the need for the knowledge of the full network gradient in the learning rule and by requiring the output network error as the only non-local information. In the augmented DFA method presented in \cite{nakajima2022physical}, the network knowledge required by the learning rule is further reduced, by replacing the differential of the activation function with an arbitrary nonlinear function. This results in a hardware friendly deep learning approach approximating BP, demonstrated both with software examples and within optoelectronic hardware (deep reservoir computer). Interestingly, high performance is obtained for different benchmark tasks. Importantly, the PPRRNN here proposed can be in principle trained employing this augmented DFA approach, by using the input signal to convey the output overall error and thus letting the plastic weights adapt to the error information.

\section{Conclusion}
We presented an experimental investigation of a new type of integrated photonic artificial neural network based on silicon ring resonators and phase change material cells (GST). We demonstrate, for the first time, complex nonlinear behaviour and multi-scale volatile memory (provided by silicon nonlinear effects), concurrently with all-optical non-volatile memory (provided by GST cells).

We investigated how our network can plastically adapt to different input temporal sequences, thanks to the non-volatile all-optical memory introduced by the phase change material cells. This adaptation happens in an emergent way, and does not rely on external control. As part of this study on plasticity mechanisms, we investigate a simple but highly nonlinear machine learning problem, consisting of the classification of 5 different temporal sequences of 4 optical pulses. We applied a novel method to modify the network internal weights exclusively via different input signals (leveraging plastic adaptation) and we showed that these modifications often significantly improve the machine learning performances compared to initial configurations. 

Moreover, in order to evaluate how powerful is the presented system in practice, we tackled a benchmark machine learning task, namely the classification of images from the MNIST dataset. Each image was inserted in the photonic network as a temporal sequence. The employed ML model does not require backprogation and consists in combining several linear classifiers (through the \textit{chaining} ensemble method) applied to different parallel outputs of our neuromorphic hardware, where each output provides a different nonlinear representation of the input image. We achieved a surprisingly high maximum average accuracy of 98.2\% and we compared it with the results from other recent works about experimental neuromorphic computing with photonics.

Finally, we discussed some properties and mechanisms enabling scalability of the proposed photonic integrated network compared to other more conventional neuromorphic computing systems, designed to be trained externally, usually via BP. 

These results lay the groundwork for the application of biologically plausible and hardware-friendly training approaches (potentially inspired by e.g. \cite{hinton2022forward} and \cite{nakajima2022physical}), by exploiting the emergent plasticity property and thus without explicitly tuning the network weights. This type of training approach is particularly interesting for neuromorphic computing research, since biological brains learn and memorize by means of plastic adaptation. This allows to forego additional external connections used to tune the network parameters, as is, instead, usually required by the training of today's neuromorphic systems. Such a possibility could greatly increase the scalability of hardware ANNs, since it would allow to employ extremely complex physical systems with a large number of nodes and plastic connections, without the need to externally access every synaptic weight and neuron.

\section{Methods}

\subsection{Experimental setup}
\begin{figure}[!htbp]
  \includegraphics[width=350pt]{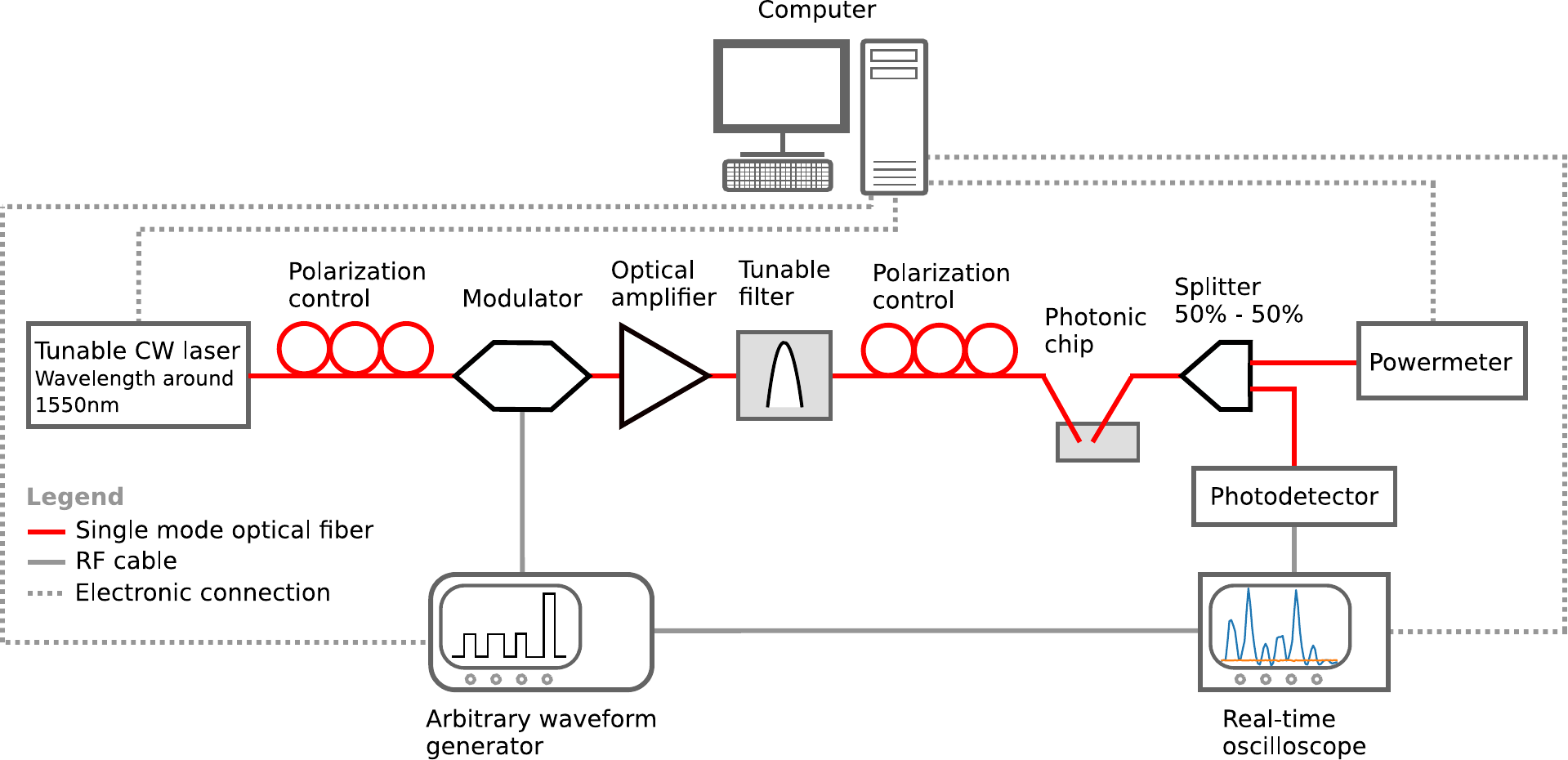}
  \centering
  \caption{\textbf{Experimental setup}.}
  \label{fig:setup}
\end{figure}

We employed a setup (see Fig. \ref{fig:setup}) capable of generating a time-dependent optical signal (max. bandwidth around 300 MHz), inserting it into a photonic integrated circuit and acquiring the response. The output of a Santec TSL-550 tunbale laser was modulated by an X-blue 40 GHz modulator controlled by an arbitrary waveform generator (AWG) (Moku:Lab). The signal was then amplified by an EDFA (Keopsys) and filtered with a band-pass filter centered on the laser wavelength. The clean and modulated optical signal was coupled into and out of the photonic chip by means of fiber grating couplers. The output of the integrated circuit was split so that roughly half of the power would reach a power meter measuring the average light power, used to estimate the optical power coupled into the chip. The other output of the splitter was measured by a fast photodetector (Thorlabs balanced photodetector 1.6 GHz), whose RF electric output was acquired by an oscilloscope (Keysight Infiniivision 3000 X-Series). A Python algorithm was running on a PC to synchronize the operations of the tunable laser (controlling power and wavelength), the AWG (controlling the type of generated waveforms), the oscilloscope and the power meter (used as reference to calculate the on-chip optical power).

\subsection{Design and fabrication} 
The photonic circuitry was fabricated through e-beam technology using shallow-etched waveguides. The considered MRRs have a radius of $\SI{15}{\micro\meter}$, a coupling gap of $\SI{350}{\nano\meter}$ and a GST patch covering a section of the ring resonator of $\SI{1}{\micro\meter}$ long.
On the same chip, within around 63 mm\(^2\), we designed and fabricated 120 PPRRNNs with different topologies, number of nodes, sparsity of GST cells and coupling gap of the MRRs.

\subsection{Practical details of plasticity investigation} \label{subsec:methods_plasticityInvestigation}
With reference to Section \ref{sec:plasticity}, by alternating inference and PA steps, we investigated how different classes of pump waveforms could achieve different non-volatile weight configurations, and how this would impact the ML performance. In particular, once we chose an input port in the investigated PPRRNN, we looked for a wavelength range where significant waveform distortions would appear at the direct output (i.e. the output port that is directly connected to the input port by a straight waveguide). Then, we performed the first inference step, changing the input wavelength with 21 steps of $\SI{0.005}{\nano\meter}$, thus spanning over a total range of $\SI{0.1}{\nano\meter}$. This measurement was repeated for each output port and resulted in a well-readable signal, each time acquiring between 70 and 80 waveforms per class. Therefore, an inference step provides 21 ML datasets, one for each wavelength, used to train and test the logistic regression. Moreover, we have as many ML features as the number of measured output ports. The first inference step provides a performance estimation of the considered network part (given by input port and wavelength range) with the initial non-volatile weight configuration. Afterwards, a subsequent PA step (usually of the first waveform class) modifies the non-volatile weights configuration, which is then evaluated through a second inference step, and so on. Additionally, we employ two nested cross-validation loops: the inner one to optimize the L2 penalty, the outer to test the trained model on all the available data.

The plotted data in Fig. \ref{fig:plasticity}a was obtained using the third input port on the left (with reference to Fig. \ref{fig:plasticity}c) and reading the signals at the output ports number 1, 2, 3, 4, 6, 8 on the right. The complete list of performed measurement sessions can be found in Table \ref{tab:meas_sess}.

\rowcolors{2}{gray!15}{white}
\begin{table}
\caption{List of performed measurements considered for the plasticity investigation presented in Section \ref{sec:plasticity}.}
  \centering
  \begin{tabular}{p{3cm}p{1.4cm}p{1.8cm}p{3.5cm}p{3.5cm}}
    \toprule
    Measurement session & Input port & Output ports & Classes in PA steps & Wavelength range center \\
    \midrule
   1 & 3 & 1,3,5,9 & 2,3,4,5 & $\SI{1547.06}{\nano\meter}$ \\
2 & 3 & 1,3,5,7 & 5 & $\SI{1547.16}{\nano\meter}$ \\ 
3 & 2 & 1,2,3,4,6 & 1,2,3,5,1 & $\SI{1548.28}{\nano\meter}$ \\ 
4 & 1 & 1 & 1,2,3,5,1,2,3  & $\SI{1547.85}{\nano\meter}$ \\ 
5 & 2 & 2 & 6 repetitions of 1,2,3,4,5 & $\SI{1549.27}{\nano\meter}$ (single)\\
6 & 2 & 2 & 6 repetitions of 1,2,3,4,5 & $\SI{1548.12}{\nano\meter}$ (single)\\
7 & 2 & 2 & 6 repetitions of 1,2,3,4,5 & $\SI{1547.86}{\nano\meter}$ (single)\\
8 & 2 & 2 & 6 repetitions of 1,2,3,4,5 & $\SI{1547.28}{\nano\meter}$ (single)\\
9 & 2 & 2 & 6 repetitions of 1,2,3,4,5 & $\SI{1547.26}{\nano\meter}$ (single)\\
10 & 2 & 2 & 6 repetitions of 1,2,3,4,5 & $\SI{1547.19}{\nano\meter}$ (single)\\
11 & 2 & 2 & 6 repetitions of 1,2,3,4,5 & $\SI{1547.17}{\nano\meter}$ (single)\\
12 & 1 & 1 & 6 repetitions of 1,2,3,4,5 & $\SI{1548.28}{\nano\meter}$ (single)\\
    \bottomrule
  \end{tabular}
  \label{tab:meas_sess}
\end{table}

\subsection{Machine learning and measurement aspects} \label{sec:methods_MLaspects}
Regarding the MNIST classification task described in Section \ref{sec:MNIST}, the presented results are obtained considering 12 input-output configurations in the network shown in Fig. \ref{fig:MNIST}a:
\begin{itemize}
\item 3 representations using the $5^{\text{th}}$ input port (counting the left grating couplers from top to bottom) and input wavelength $\lambda=\SI{1548.14}{\nano\meter}$, at $3^\text{rd}$, $5^\text{th}$ and $13^\text{th}$ output ports (right grating couplers from top to bottom).
\item  5 representations using the $6^{\text{th}}$ input port with $\lambda = \SI{1547.27}{\nano\meter}$, at output port numbers 4, 6, 8, 10, 12.
\item  4 representations using again the $6^{\text{th}}$ input port but with $\lambda = \SI{1548.77}{\nano\meter}$, at output port numbers 2, 4, 6, 8.
\end{itemize}
Because of limitations exclusively due to instrumentation (memory and speed of electronics to generate input waveforms and to acquire output waveforms), we employed a subset of 2941 images with balanced classes. Each image was inserted into the PPRRNN 4 or 5 times, thus performing data augmentation to improve the learning of experimental noise by the training algorithm, reaching a total sample number of 13466. Due to the same setup limitations, we downsampled the images from 28$\times$28 to 14$\times$14 pixels, employing the maximum over adjacent squares composed of four neighbouring pixels (Fig. \ref{fig:MNIST}a), in order to reduce the information loss due to downsampling. Each image was then flattened to a 1D array, upsampled with a factor of 2 using linear interpolation and inserted into the PPRRNN as a single waveform. It should be stressed that the preprocessing is not meant to make the ML task easier (on the contrary, it probably makes it harder because of loss of information due to downsampling) and, apart from the flattening, it should be avoided if the instrumentation allows. We employed a bit duration of $\SI{4}{\nano\second}$ and each flattened image was temporally separated from the next by a no-signal period of around $\SI{2}{\micro\second}$, in order to avoid temporal cross-talk due to thermo-optic effects.

We encoded the MNIST class label (i.e. the ground truth) corresponding to each output waveform in the time distance between the current waveform and the next generated one. This way we could retrieve the class labels directly from the output of the PPRRNN, while ensuring that the label information could not be retrieved by the readout linear classifiers. In particular, the time distance between each generated input waveform was set to $\SI{2}{\micro\second} + \SI{24}{\nano\second}\: \times$ class label (from 0 to 9, labelling the previous waveform). The output waveforms are extracted from the acquired data by detecting the presence of a signal above the noise floor, which usually corresponds to the beginning of the insertion of the input waveform, since the light delay is negligible. In those cases where this was not true, because of low transmission of the full waveform or of its initial part through the network, the output waveform was discarded.

In Section \ref{sec:MNIST} we present the results obtained considering the 12 input-output configurations in the network shown in Fig. \ref{fig:MNIST}a. Let us first explain our ML pipeline (Fig. \ref{fig:MNIST}b), which consists of building a combination (chain) of linear classifiers, each applied to a different output representation. The chain is automatically built by a greedy algorithm, which adds one classifier to the chain at each step, trying to correct the mistakes made by the previous classifiers. Assume we measure \(N\) output nonlinear representations for each input image. Then, we split the available ML dataset into \(N+2\) folds with an approximately equal number of samples per class. We keep the last fold apart to use it at the very end for testing. In order to select the best stand-alone output representation, which will be the first in our hierarchical chain, we apply an individual linear classifier to the samples of each output representation and train it on fold 1. In fact, this is analogous to training \(N\) reservoir computing (RC) systems \cite{tanaka2019recent,lugnan2020photonic}, since we can consider each representation as the outcome of a different untrained recurrent neural network with \(n\) temporal outputs. Subsequently, we select the best performing classifier, by evaluation of a validation set containing folds 2 to \(N+1\) in the dataset. In general, each single output classifier could not outperform the baseline of directly applying a linear classifier to the input waveform (\(\sim 86\%\) accuracy in our experiment, \(88\%\) in software \cite{lecun1998gradient}). This is indeed expected, because each representation is neither optimized to improve accuracy, nor has a sufficiently high dimensionality compared to the input, which is required for traditional RC.

In the second step of our pipeline, we build an ensemble of two classifiers by employing the \textit{chaining} method \cite{freiberger2019improving}. Recall that the output of the first classifier consists of 10 numbers, each being the estimated probability of belonging to one of the classes. Then, we train a second classifier on these 10 numbers combined with the samples of another representation from a different (virtual) network (Fig. \ref{fig:MNIST}b). We train this linear classifier on folds 1 and 2 of the dataset and we validate its performance on folds from 3 to \(N+1\). This way, the second linear classifier in the chain focuses especially on correcting the errors made by the first classifier (which was trained on fold 1) in generalizing over fold 2. Therefore, this method aims to progressively improve the computational power of the ensemble of the classifiers, while reducing overfitting. Among all the \(N-1\) possible 2-reservoirs chaining combinations, we select the one with highest validation accuracy. 

Afterwards, starting from the selected chain of two classifiers, we repeat the process so as to select the best 3-classifier chain, and so on until we obtain a chain of length \(N\), trained on folds from 1 to \(N\) and validated on fold \(N+1\). We employ the resulting validation score to optimize the regularization strength of the L2 penalty in the training of the last linear classifier in the chain. Finally, we estimate the test accuracy of the obtained \(N\)-reservoir chain using the unseen fold \(N+2\). We repeat this whole chain formation \(N+2\) times, each time using a different fold to select the first classifier, so as to perform a \textit{k-fold cross-validation}, where \(k = N+2\). This way, we evaluate our ML pipeline on all the available data, in order to maximize the precision of our test accuracy evaluation.

In our experiments, the output representations at different wavelengths and physical ports are acquired sequentially, one after another. However, we consider these results to be a good approximation of a truly parallel measurement, where many photodetectors are employed at the same time, together with filters to separate the different wavelengths. Indeed, the three considered wavelengths are distant enough from each others so that they cannot be significantly coupled by the nonlinear response of the MRRs. Moreover, since the sample insertion is repeated a large number of times (much larger even than the measured repetitions) we believe that plastic changes have reached an equilibrium by the time they are measured, allowing repeatability. Indeed, if significant changes over time were to occur during the repetitions, the classification task presented to the linear classifier would artificially become more difficult to carry out, thus limiting the achievable accuracy.

\subsection{Energy, footprint and throughput of the proposed integrated photonic network} \label{sec:methods_energyConsumption}
In this work, we employed a peak on-chip power (for a single input wavelength) of around $\SI{21}{\milli\watt}$ for a fully white input pixels, and a   power of around $\SI{1}{\milli\watt}$ for a black pixel. This corresponds to an on-chip input energy per white and black pixel respectively of around $\SI{84}{\pico\joule}$ and $\SI{4}{\pico\joule}$. An upper estimate for the average on-chip energy per image is $\SI{17}{\nano\joule}$, which is found by assuming half the pixels to be completely white and the other half completely black: \( \SI{17}{\nano\joule} \approx  196 \times \SI{84}{\pico\joule}+ 196 \times \SI{4}{\pico\joule}\). 
Regarding the on-chip footprint, the PPRRNN considered in this section takes up around $\SI{0.5}{\milli\meter}^2$, providing 7 physical output ports on the right side and 6 on the left side (even though the number of ports is doubled if the counterpropagating field is strong enough to be readable). Therefore, we can estimate that our PPRRNN can potentially provide at least 13 nonlinear representations per mm$^2$ per wavelength. A large number of representations (hundreds or even thousands) could be generated on a single chip by employing several wavelengths at the same time and considering larger or multiple PPRRNNs. In practice though, one should consider the feasibility and the impact of having many input and output optical connections on the same chip, of separating several wavelengths at the output, of employing a large number of photodetectors, of managing thermal cross-talk, etc. However, being able to generate a large number of representations on a small chip area can be advantageous even if the representations are not read out all at the same time. Indeed, the achievable representations could be explored by automatic measurements even one by one, so as to select the best few.

Finally, here we explain how we estimated the approximated aerial and energy efficiency of a PPRRNN in terms of multiply-accumulate operations (MACs) plus nonlinear operations (NLOs), namely \(10^{15}\) (MACs+NLOs)/s/mm\(^2\) and \(5\times10^{15}\) (MACs+NLOs)/J, which are reported in Section \ref{sec:MNIST}. First, as we did for the demonstrations presented in this paper, let us consider the use of free-carrier based nonlinearity to achieve the activation function of the photonic neurons, and the temperature-based nonlinearity as volatile memory which can integrate several neuron activations through time. A realistic case is that we use \SI{2}{\nano \second} input temporal resolution, a time duration sufficient for free-carriers to provide strong nonlinearity, and an integration time due to the thermal memory of around \SI{200}{\nano \second}. Therefore, considering only one input port connected to one output port in a PPRRNN, we have that each \SI{2}{\nano \second} at the output is the result of a nonlinear integration of the input inserted in the previous \SI{200}{\nano \second}. Thus 100 MACs+NLOs operations are performed each \SI{2}{\nano \second}, considering only the time dimension. Equivalently, we can see this system as 2 connected neuron layers (input and output) in the time domain, each comprising 100 neurons, neglecting for simplicity the recurrent operations in the hidden node layers. The number of MACs+NLOs is found by multiplying the dimension of the two layers, which gives \(10^4\) (MACs+NLOs) each \SI{200}{\nano \second}. Moreover, in a PPRRNN fitting a 0.5 mm\(^2\) chip area, it is realistic to have at least 10 physical input ports connected to other 10 output physical ports, hence we obtain an input and an output layers of \(10^3\) neurons each, achieving \(10^6\) MACs+NLOs per \SI{200}{\nano \second} and per 0.5 mm\(^2\), covering both the time and spatial dimensions. Furthermore, in such a PPRRNN we can in principle employ at least 10 input wavelength channels per physical port, so that they are interconnected by the network activity, thus further expanding both the input and output dimensions from \(10^3\)  to \(10^4\) each, achieving a total throughput per unit area of \(10^{15}\)(MACs+NLOs)/s/mm\(^2\). Regarding the energy efficiency, we consider a realistic input power of \SI{10}{\milli \watt} per input physical port, yielding a power consumption of around \SI{200}{\milli \watt}/mm\(^2\). Dividing the throughput per area by this quantity, we finally find a power efficiency of \(5\times10^{15}\) (MACs+NLOs)/J.

\section*{Data availability}
The datasets generated during the current study are available from the corresponding author on reasonable request.

\bibliographystyle{unsrt}  
\bibliography{references}  

\begin{thebibliography}{10}

\bibitem{lecun2015deep}
Yann LeCun, Yoshua Bengio, and Geoffrey Hinton.
\newblock Deep learning.
\newblock {\em nature}, 521(7553):436--444, 2015.

\bibitem{dong2021survey}
Shi Dong, Ping Wang, and Khushnood Abbas.
\newblock A survey on deep learning and its applications.
\newblock {\em Computer Science Review}, 40:100379, 2021.

\bibitem{strubell2019energy}
Emma Strubell, Ananya Ganesh, and Andrew McCallum.
\newblock Energy and policy considerations for deep learning in nlp.
\newblock {\em arXiv preprint arXiv:1906.02243}, 2019.

\bibitem{thompson2020computational}
Neil~C Thompson, Kristjan Greenewald, Keeheon Lee, and Gabriel~F Manso.
\newblock The computational limits of deep learning.
\newblock {\em arXiv preprint arXiv:2007.05558}, 2020.

\bibitem{wu2022sustainable}
Carole-Jean Wu, Ramya Raghavendra, Udit Gupta, Bilge Acun, Newsha Ardalani,
  Kiwan Maeng, Gloria Chang, Fiona Aga, Jinshi Huang, Charles Bai, et~al.
\newblock Sustainable ai: Environmental implications, challenges and
  opportunities.
\newblock {\em Proceedings of Machine Learning and Systems}, 4:795--813, 2022.

\bibitem{schmidgall2023brain}
Samuel Schmidgall, Jascha Achterberg, Thomas Miconi, Louis Kirsch, Rojin Ziaei,
  S~Hajiseyedrazi, and Jason Eshraghian.
\newblock Brain-inspired learning in artificial neural networks: a review.
\newblock {\em arXiv preprint arXiv:2305.11252}, 2023.

\bibitem{lillicrap2020backpropagation}
Timothy~P Lillicrap, Adam Santoro, Luke Marris, Colin~J Akerman, and Geoffrey
  Hinton.
\newblock Backpropagation and the brain.
\newblock {\em Nature Reviews Neuroscience}, 21(6):335--346, 2020.

\bibitem{taherkhani2020review}
Aboozar Taherkhani, Ammar Belatreche, Yuhua Li, Georgina Cosma, Liam~P Maguire,
  and T~Martin McGinnity.
\newblock A review of learning in biologically plausible spiking neural
  networks.
\newblock {\em Neural Networks}, 122:253--272, 2020.

\bibitem{jeon2023distinctive}
Ikhwan Jeon and Taegon Kim.
\newblock Distinctive properties of biological neural networks and recent
  advances in bottom-up approaches toward a better biologically plausible
  neural network.
\newblock {\em Frontiers in Computational Neuroscience}, 17, 2023.

\bibitem{hinton2022forward}
Geoffrey Hinton.
\newblock The forward-forward algorithm: Some preliminary investigations.
\newblock {\em arXiv preprint arXiv:2212.13345}, 2022.

\bibitem{nakajima2022physical}
Mitsumasa Nakajima, Katsuma Inoue, Kenji Tanaka, Yasuo Kuniyoshi, Toshikazu
  Hashimoto, and Kohei Nakajima.
\newblock Physical deep learning with biologically inspired training method:
  gradient-free approach for physical hardware.
\newblock {\em Nature Communications}, 13(1):7847, 2022.

\bibitem{magee2020synaptic}
Jeffrey~C Magee and Christine Grienberger.
\newblock Synaptic plasticity forms and functions.
\newblock {\em Annual review of neuroscience}, 43:95--117, 2020.

\bibitem{xu2023reconfigurable}
Minyi Xu, Xinrui Chen, Yehao Guo, Yang Wang, Dong Qiu, Xinchuan Du, Yi~Cui,
  Xianfu Wang, and Jie Xiong.
\newblock Reconfigurable neuromorphic computing: Materials, devices and
  integration.
\newblock {\em Advanced Materials}, page 2301063, 2023.

\bibitem{lockwood2021silicon}
David~J Lockwood and Lorenzo Pavesi.
\newblock {\em Silicon Photonics IV}, volume 139.
\newblock Springer, 2021.

\bibitem{markovic2020physics}
Danijela Markovi{\'c}, Alice Mizrahi, Damien Querlioz, and Julie Grollier.
\newblock Physics for neuromorphic computing.
\newblock {\em Nature Reviews Physics}, 2(9):499--510, 2020.

\bibitem{schuman2022opportunities}
Catherine~D Schuman, Shruti~R Kulkarni, Maryam Parsa, J~Parker Mitchell,
  Prasanna Date, and Bill Kay.
\newblock Opportunities for neuromorphic computing algorithms and applications.
\newblock {\em Nature Computational Science}, 2(1):10--19, 2022.

\bibitem{christensen20222022}
Dennis~V Christensen, Regina Dittmann, Bernabe Linares-Barranco, Abu Sebastian,
  Manuel Le~Gallo, Andrea Redaelli, Stefan Slesazeck, Thomas Mikolajick, Sabina
  Spiga, Stephan Menzel, et~al.
\newblock 2022 roadmap on neuromorphic computing and engineering.
\newblock {\em Neuromorphic Computing and Engineering}, 2(2):022501, 2022.

\bibitem{shastri2021photonics}
Bhavin~J Shastri, Alexander~N Tait, Thomas Ferreira~de Lima, Wolfram~HP
  Pernice, Harish Bhaskaran, C~David Wright, and Paul~R Prucnal.
\newblock Photonics for artificial intelligence and neuromorphic computing.
\newblock {\em Nature Photonics}, 15(2):102--114, 2021.

\bibitem{pavanello2023special}
Fabio Pavanello, Elena~Ioana Vatajelu, Alberto Bosio, Thomas Van~Vaerenbergh,
  Peter Bienstman, Benoit Charbonnier, Alessio Carpegna, Stefano Di~Carlo, and
  Alessandro Savino.
\newblock Special session: Neuromorphic hardware design and reliability from
  traditional cmos to emerging technologies.
\newblock In {\em 2023 IEEE 41st VLSI Test Symposium (VTS)}, pages 1--10. IEEE,
  2023.

\bibitem{wuttig2017phase}
Matthias Wuttig, Harish Bhaskaran, and Thomas Taubner.
\newblock Phase-change materials for non-volatile photonic applications.
\newblock {\em Nature photonics}, 11(8):465--476, 2017.

\bibitem{feldmann2019all}
Johannes Feldmann, Nathan Youngblood, C~David Wright, Harish Bhaskaran, and
  Wolfram~HP Pernice.
\newblock All-optical spiking neurosynaptic networks with self-learning
  capabilities.
\newblock {\em Nature}, 569(7755):208--214, 2019.

\bibitem{tait2019silicon}
Alexander~N Tait, Thomas~Ferreira De~Lima, Mitchell~A Nahmias, Heidi~B Miller,
  Hsuan-Tung Peng, Bhavin~J Shastri, and Paul~R Prucnal.
\newblock Silicon photonic modulator neuron.
\newblock {\em Physical Review Applied}, 11(6):064043, 2019.

\bibitem{amin2019ito}
Rubab Amin, JK~George, Shuai Sun, Thomas Ferreira~de Lima, Alexander~N Tait,
  JB~Khurgin, Mario Miscuglio, Bhavin~J Shastri, Paul~R Prucnal, Tarek
  El-Ghazawi, et~al.
\newblock Ito-based electro-absorption modulator for photonic neural activation
  function.
\newblock {\em APL Materials}, 7(8), 2019.

\bibitem{nahmias2016integrated}
Mitchell~A Nahmias, Alexander~N Tait, Leonidas Tolias, Matthew~P Chang, Thomas
  Ferreira~de Lima, Bhavin~J Shastri, and Paul~R Prucnal.
\newblock An integrated analog o/e/o link for multi-channel laser neurons.
\newblock {\em Applied Physics Letters}, 108(15), 2016.

\bibitem{buckley2023photonic}
Sonia~Mary Buckley, Alexander~N Tait, Adam~N McCaughan, and Bhavin~J Shastri.
\newblock Photonic online learning: a perspective.
\newblock {\em Nanophotonics}, 12(5):833--845, 2023.

\bibitem{pai2022experimentally}
Sunil Pai, Zhanghao Sun, Tyler~W Hughes, Taewon Park, Ben Bartlett, Ian~AD
  Williamson, Momchil Minkov, Maziyar Milanizadeh, Nathnael Abebe, Francesco
  Morichetti, et~al.
\newblock Experimentally realized in situ backpropagation for deep learning in
  nanophotonic neural networks.
\newblock {\em arXiv preprint arXiv:2205.08501}, 2022.

\bibitem{bandyopadhyay2022single}
Saumil Bandyopadhyay, Alexander Sludds, Stefan Krastanov, Ryan Hamerly,
  Nicholas Harris, Darius Bunandar, Matthew Streshinsky, Michael Hochberg, and
  Dirk Englund.
\newblock Single chip photonic deep neural network with accelerated training.
\newblock {\em arXiv preprint arXiv:2208.01623}, 2022.

\bibitem{lecun2010MNIST}
Yann LeCun and Corinna Cortes.
\newblock {MNIST} handwritten digit database.
\newblock http://yann.lecun.com/exdb/mnist/, 2010.

\bibitem{bogaerts2012silicon}
Wim Bogaerts, Peter De~Heyn, Thomas Van~Vaerenbergh, Katrien De~Vos, Shankar
  Kumar~Selvaraja, Tom Claes, Pieter Dumon, Peter Bienstman, Dries
  Van~Thourhout, and Roel Baets.
\newblock Silicon microring resonators.
\newblock {\em Laser \& Photonics Reviews}, 6(1):47--73, 2012.

\bibitem{van2012cascadable}
Thomas Van~Vaerenbergh, Martin Fiers, Pauline Mechet, Thijs Spuesens, Rajesh
  Kumar, Geert Morthier, Benjamin Schrauwen, Joni Dambre, and Peter Bienstman.
\newblock Cascadable excitability in microrings.
\newblock {\em Optics express}, 20(18):20292--20308, 2012.

\bibitem{mesaritakis2013micro}
Charis Mesaritakis, Vassilis Papataxiarhis, and Dimitris Syvridis.
\newblock Micro ring resonators as building blocks for an all-optical
  high-speed reservoir-computing bit-pattern-recognition system.
\newblock {\em JOSA B}, 30(11):3048--3055, 2013.

\bibitem{mancinelli2014chaotic}
M~Mancinelli, M~Borghi, F~Ramiro-Manzano, JM~Fedeli, and L~Pavesi.
\newblock Chaotic dynamics in coupled resonator sequences.
\newblock {\em Optics express}, 22(12):14505--14516, 2014.

\bibitem{lugnan2022rigorous}
Alessio Lugnan, Santiago Garc{\'\i}a-Cuevas Carrillo, C~David Wright, and Peter
  Bienstman.
\newblock Rigorous dynamic model of a silicon ring resonator with phase change
  material for a neuromorphic node.
\newblock {\em Optics Express}, 30(14):25177--25194, 2022.

\bibitem{biasi2023photonic}
Stefano Biasi, Giovanni Donati, Alessio Lugnan, Mattia Mancinelli, Emiliano
  Staffoli, and Lorenzo Pavesi.
\newblock Photonic neural networks based on integrated silicon microresonators.
\newblock {\em arXiv preprint arXiv:2306.04779}, 2023.

\bibitem{chakraborty2018toward}
Indranil Chakraborty, Gobinda Saha, Abhronil Sengupta, and Kaushik Roy.
\newblock Toward fast neural computing using all-photonic phase change spiking
  neurons.
\newblock {\em Scientific reports}, 8(1):12980, 2018.

\bibitem{feldmann2021parallel}
Johannes Feldmann, Nathan Youngblood, Maxim Karpov, Helge Gehring, Xuan Li,
  Maik Stappers, Manuel Le~Gallo, Xin Fu, Anton Lukashchuk, Arslan~Sajid Raja,
  et~al.
\newblock Parallel convolutional processing using an integrated photonic tensor
  core.
\newblock {\em Nature}, 589(7840):52--58, 2021.

\bibitem{tanaka2019recent}
Gouhei Tanaka, Toshiyuki Yamane, Jean~Benoit H{\'e}roux, Ryosho Nakane, Naoki
  Kanazawa, Seiji Takeda, Hidetoshi Numata, Daiju Nakano, and Akira Hirose.
\newblock Recent advances in physical reservoir computing: A review.
\newblock {\em Neural Networks}, 115:100--123, 2019.

\bibitem{lugnan2020photonic}
Alessio Lugnan, Andrew Katumba, Floris Laporte, Matthias Freiberger, Stijn
  Sackesyn, Chonghuai Ma, Emmanuel Gooskens, Joni Dambre, and Peter Bienstman.
\newblock Photonic neuromorphic information processing and reservoir computing.
\newblock {\em APL Photonics}, 5(2), 2020.

\bibitem{dabos2022neuromorphic}
George Dabos, Dimitris~V Bellas, Ripalta Stabile, Miltiadis Moralis-Pegios,
  George Giamougiannis, Apostolos Tsakyridis, Angelina Totovic, Elefterios
  Lidorikis, and Nikos Pleros.
\newblock Neuromorphic photonic technologies and architectures: scaling
  opportunities and performance frontiers.
\newblock {\em Optical Materials Express}, 12(6):2343--2367, 2022.

\bibitem{mourgias2022noise}
G~Mourgias-Alexandris, M~Moralis-Pegios, A~Tsakyridis, S~Simos, G~Dabos,
  A~Totovic, N~Passalis, M~Kirtas, T~Rutirawut, FY~Gardes, et~al.
\newblock Noise-resilient and high-speed deep learning with coherent silicon
  photonics.
\newblock {\em Nature Communications}, 13(1):5572, 2022.

\bibitem{antonik2019large}
Piotr Antonik, Nicolas Marsal, and Damien Rontani.
\newblock Large-scale spatiotemporal photonic reservoir computer for image
  classification.
\newblock {\em IEEE Journal of Selected Topics in Quantum Electronics},
  26(1):1--12, 2019.

\bibitem{nakajima2021scalable}
Mitsumasa Nakajima, Kenji Tanaka, and Toshikazu Hashimoto.
\newblock Scalable reservoir computing on coherent linear photonic processor.
\newblock {\em Communications Physics}, 4(1):20, 2021.

\bibitem{bai2023microcomb}
Bowen Bai, Qipeng Yang, Haowen Shu, Lin Chang, Fenghe Yang, Bitao Shen, Zihan
  Tao, Jing Wang, Shaofu Xu, Weiqiang Xie, et~al.
\newblock Microcomb-based integrated photonic processing unit.
\newblock {\em Nature Communications}, 14(1):66, 2023.

\bibitem{zhu2022space}
HH~Zhu, Jun Zou, Hengyi Zhang, YZ~Shi, SB~Luo, N~Wang, H~Cai, LX~Wan, Bo~Wang,
  XD~Jiang, et~al.
\newblock Space-efficient optical computing with an integrated chip diffractive
  neural network.
\newblock {\em Nature communications}, 13(1):1044, 2022.

\bibitem{jaeger2021dimensions}
Herbert Jaeger, Dirk Doorakkers, Celestine Lawrence, and Giacomo Indiveri.
\newblock Dimensions of timescales in neuromorphic computing systems.
\newblock {\em arXiv preprint arXiv:2102.10648}, 2021.

\bibitem{vandoorne2014experimental}
Kristof Vandoorne, Pauline Mechet, Thomas Van~Vaerenbergh, Martin Fiers, Geert
  Morthier, David Verstraeten, Benjamin Schrauwen, Joni Dambre, and Peter
  Bienstman.
\newblock Experimental demonstration of reservoir computing on a silicon
  photonics chip.
\newblock {\em Nature communications}, 5(1):3541, 2014.

\bibitem{you200812}
Jong-Bum You, Miran Park, Jeong-Woo Park, and Gyungock Kim.
\newblock 12.5 gbps optical modulation of silicon racetrack resonator based on
  carrier-depletion in asymmetric pn diode.
\newblock {\em Optics express}, 16(22):18340--18344, 2008.

\bibitem{nokland2016direct}
Arild N{\o}kland.
\newblock Direct feedback alignment provides learning in deep neural networks.
\newblock {\em Advances in neural information processing systems}, 29, 2016.

\bibitem{lecun1998gradient}
Yann LeCun, L{\'e}on Bottou, Yoshua Bengio, and Patrick Haffner.
\newblock Gradient-based learning applied to document recognition.
\newblock {\em Proceedings of the IEEE}, 86(11):2278--2324, 1998.

\bibitem{freiberger2019improving}
Matthias Freiberger, Stijn Sackesyn, Chonghuai Ma, Andrew Katumba, Peter
  Bienstman, and Joni Dambre.
\newblock Improving time series recognition and prediction with networks and
  ensembles of passive photonic reservoirs.
\newblock {\em IEEE Journal of Selected Topics in Quantum Electronics},
  26(1):1--11, 2019.

\end{thebibliography}


\begin{thebibliography}{1}

\bibitem{lugnan2022rigorous}
Alessio Lugnan, Santiago Garc{\'\i}a-Cuevas Carrillo, C~David Wright, and Peter
  Bienstman.
\newblock Rigorous dynamic model of a silicon ring resonator with phase change
  material for a neuromorphic node.
\newblock {\em Optics Express}, 30(14):25177--25194, 2022.

\bibitem{li2020experimental}
Xuan Li, Nathan Youngblood, Zengguang Cheng, Santiago Garcia-Cuevas Carrillo,
  Emanuele Gemo, Wolfram~HP Pernice, C~David Wright, and Harish Bhaskaran.
\newblock Experimental investigation of silicon and silicon nitride platforms
  for phase-change photonic in-memory computing.
\newblock {\em Optica}, 7(3):218--225, 2020.

\end{thebibliography}

\section*{Acknowledgments}
This work was funded by the European Union’s Horizon 2020 and Horizon Europe Research and Innovation Programmes (grant 780848 Fun-COMP, grant 101017237 PHOENICS, grant 101070238 NEUROPULS), by the Flemish FWO project G006020N and by the Belgian EOS project G0H1422N. We thank Andrew Katumba, Xuan Li, Johannes Feldmann and Santiago García-Cuevas Carrillo for useful discussion and help in the design and fabrication process. 

\section*{Author contribution}
A.L. and P.B. conceived the experiment; A.L. designed the integrated circuits (with help from S.A. and F.B.P.), carried out the measurements, processed and analysed the data under the supervision of P.B.; F.B.P. fabricated the integrated circuits under the supervision of W.H.P.P.; S.A. performed the deposition of the GST material under the supervision of H.B.; C.D.W. supervised device modelling and coordinated the interinstitutional collaboration; All authors wrote the manuscript together.

\section*{Competing interests}
The authors declare no competing interests.

\section*{Materials \& Correspondence}
Correspondence and material requests should be addressed to A. Lugnan (alessio.lugnan.1@unitn.it) or to P. Bienstman (peter.bienstman@ugent.be).

\end{document}


\maketitle
\section{Characterization of a single plastic node} \label{subsec:methods_singleNode}
\begin{figure}[!htbp]
  \includegraphics[width=440pt]{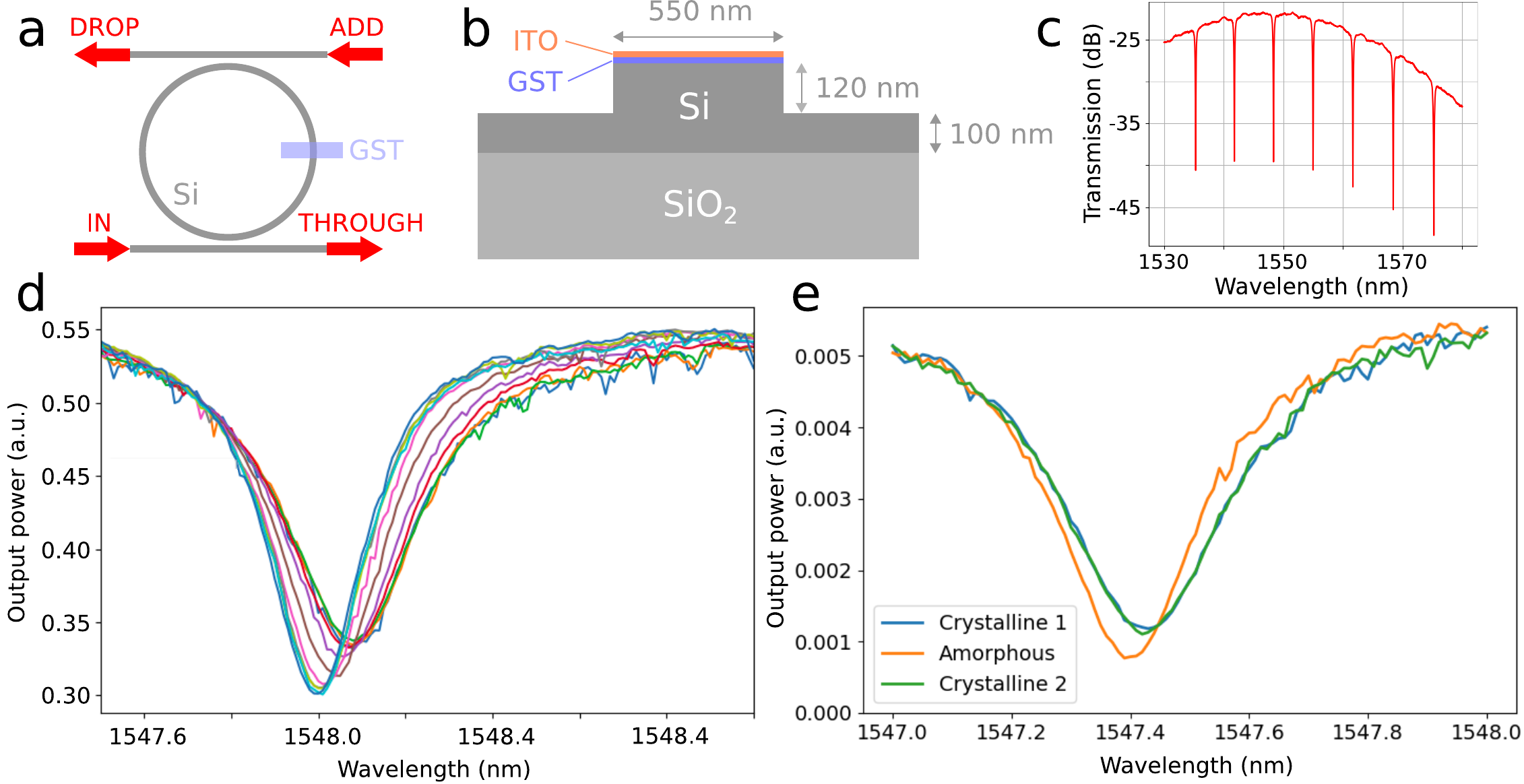}
  \centering
  \caption{\textbf{All-optical plastic node. a} The considered plastic node, consisting of a silicon add-drop MRR with a PCM cell of $\SI{1}{\micro\meter}$ \textbf{b}  PCM cell cross section. \textbf{c} Example of transmission spectrum of a silicon MRR (\textit{through} port) showing multiple quasi-periodic resonance dips, measured using grating couplers. \textbf{d} Linear response of a plastic node (RR spectra showing resonance dips at the \textit{through} output port) for different GST crystallinity levels, obtained by insertion of a single (amorphization) or multiple (crystallization) optical pulses. \textbf{e} Resonance spectrum of a plastic node, showing the non-volatile effect of single-pulse amorphization (from “Crystalline 1” to “Amorphous”) and single-waveform recrystallization (from “Amorphous” to “Crystalline 2”).}
  \label{fig:singleNode}
\end{figure}

In this supplementary section we present the experimental characterization of an integrated plastic node, namely a silicon microring resonator (MRR) with phase change material (PCM) cell (see Fig. \ref{fig:singleNode} a, b), which was used as a building block in photonic plastic networks discussed in the main text. A MRR is resonant only to wavelengths that can fit an integer number of times along the ring waveguide, so that the light travelling in the ring interferes constructively with itself after a round trip. If this condition is met, the input light accumulates in the ring and part of it is redirected to the \textit{drop} port. Otherwise, for wavelengths far from the resonant condition, most of the input signal continues its travel along the straight waveguide (to the \textit{through} port), approximately undisturbed by the presence of the resonator (see the MRR transmission spectrum in Fig.~\ref{fig:singleNode} c).
Generally, with reference to Fig.~\ref{fig:singleNode} a, depending on the wavelength of an input optical signal at the \textit{in} or \textit{add} port, and depending on the GST memory state (i.e. amorphization level), different fractions of the input light will be led to the \textit{through} port, redirected to the \textit{drop} port, or absorbed by the ring waveguide (which also comprises a PCM cell). The latter implies heating and increase in free carrier concentration of the resonator, with consequent volatile modification of the resonance properties, such as the resonant wavelengths. Furthermore, light absorption and consequent heating of the GST cell, if intense enough, triggers modifications to the non-volatile memory state. It should be also noticed that if light with the same wavelength enters through the two input ports at the same time, optical interference of overlapping beams will come into play as well. For more theoretical insight regarding this device, we refer to our previous work \cite{lugnan2022rigorous}.

In order to evaluate the energy efficiency and contrast of memory operations (considering both GST amorphization and recrystallization), we sent a sequence of short optical pulses into the MRR input port, so as to change the solid-state phase of the GST cell. In particular, a single optical pulse of $\SI{10}{\nano\second}$ duration was employed for GST amorphization, while for recrystallization we used a sequence of 100 pulses with the same duration and lower power, each separated by time intervals of $\SI{10}{\nano\second}$, with total length of $\SI{2}{\micro\second}$. We did so, instead of using the double-step pulse employed in \cite{li2020experimental}, in order to demonstrate that the repetition of the same pulse shape used for amorphization can reverse the memory state as well. This is key to achieve a plastic behaviour of the node, whose memory states need to be fully accessible by employing the same input signal shape.

By suitably setting the power of the optical input pulses we obtained different GST amorphization levels, which can be considered as intermediate memory states, each presenting a different resonance dip in the acquired MRR spectrum at the \textit{through} port (Fig. \ref{fig:singleNode}d). The chosen pulse wavelength is $\SI{0.06}{\nano\meter}$ larger than the resonance wavelength (corresponding to the center of the spectrum dip) when the GST is fully crystalline. In accordance to MRR theory, we ascribe the resonance dip changes in width and depth to variations of the GST cell absorption: a higher GST crystallization level implies a higher optical absorption, thus a larger width and a smaller depth of the resonance dip. Moreover, the GST solid-state phase also affects the effective refractive index of the corresponding waveguide segment, which in turns modifies the resonance wavelength. In particular, a larger crystalline fraction implies a larger effective refractive index and thus a larger resonance wavelength. Importantly, this effect further increases the achievable optical contrast due to different memory states, and therefore represents an additional advantage of employing a GST cell on a MRR rather than on a straight waveguide, whose transmission is much less affected by variations in effective refractive index. It should be stressed that we achieved a significantly higher optical contrast due to memory operations when experimentally investigating our plastic node compared to what we previously predicted through simulations \cite{lugnan2022rigorous} (probably because of inaccuracies in the material parameters used).

Let us now consider an example showing the maximum optical contrast, due to GST amorphization, that is achievable with a single pulse and is reversible using a single recrystallization waveform (comprising several pulses). In Fig. \ref{fig:singleNode} e, we plotted the resonance spectra of a plastic node, corresponding to the following memory states: initial crystalline GST state (labelled as ``Crystalline 1'' in the legend); partially amorphized GST after the insertion of an amorphization pulse of around $\SI{14}{\milli\watt}$ peak power, conveying around $\SI{14}{\nano\joule}$ of energy (labelled as ``Amorphous'' in the legend); return to initial crystalline state, after the insertion of a recrystallization waveform, consisting of pulses with around $\SI{1}{\milli\watt}$ peak power, conveying a total energy equal to around $\SI{1}{\nano\joule}$ (labelled as ``Crystalline 2'' in the legend).
In this case, a reversible optical contrast in the output power at 1547.4 nm greater than 40\% is achieved. In comparison, to achieve a reversible optical contrast of 15\% with a GST cell on a straight silicon waveguide (i.e. without the advantages of exploiting the MRR resonant behaviour), an amorphization pulse $\SI{100}{\nano\second}$ long and with $\SI{1.6}{\nano\joule}$ energy is required, and a recrystallization double-step pulse, $\SI{530}{\nano\second}$ long and with $\SI{3.6}{\nano\joule}$ energy \cite{lugnan2022rigorous}. Therefore, our plastic node shows the following improvements w.r.t. its straight waveguide counterpart: more than twice optical contrast; more than a factor 10 in amorphization energy efficiency; almost a factor 4 in recrystallization energy efficiency; a factor 10 in amorphization speed.
The considered recrystallization operation, instead, is more than 3 times slower (although we did not try to maximize the speed), because it employs a sequence of single-step pulses, instead of double-step pulses. The fact that we do not need specially constructed pulses is key for the network plasticity property howere, since we want the same input shape to be able to change the memory states of the plastic nodes in both directions, allowing for a more flexible network adaptability.

\section{Investigation of plastic self-adaptation properties}
In this supplementary section we provide an expansion of the information given in Section 3 of the main text, regarding the plasticity introduced in the proposed photonic plastic recurrent resonator neural network (PPRRNN) by the MRRs with PCM. In particular, here we present histograms that summarize important properties of the plastic response over all the measurement sessions performed (listed in Table 2 in the main text).

\begin{figure}[!htbp]
  \includegraphics[width=470pt]{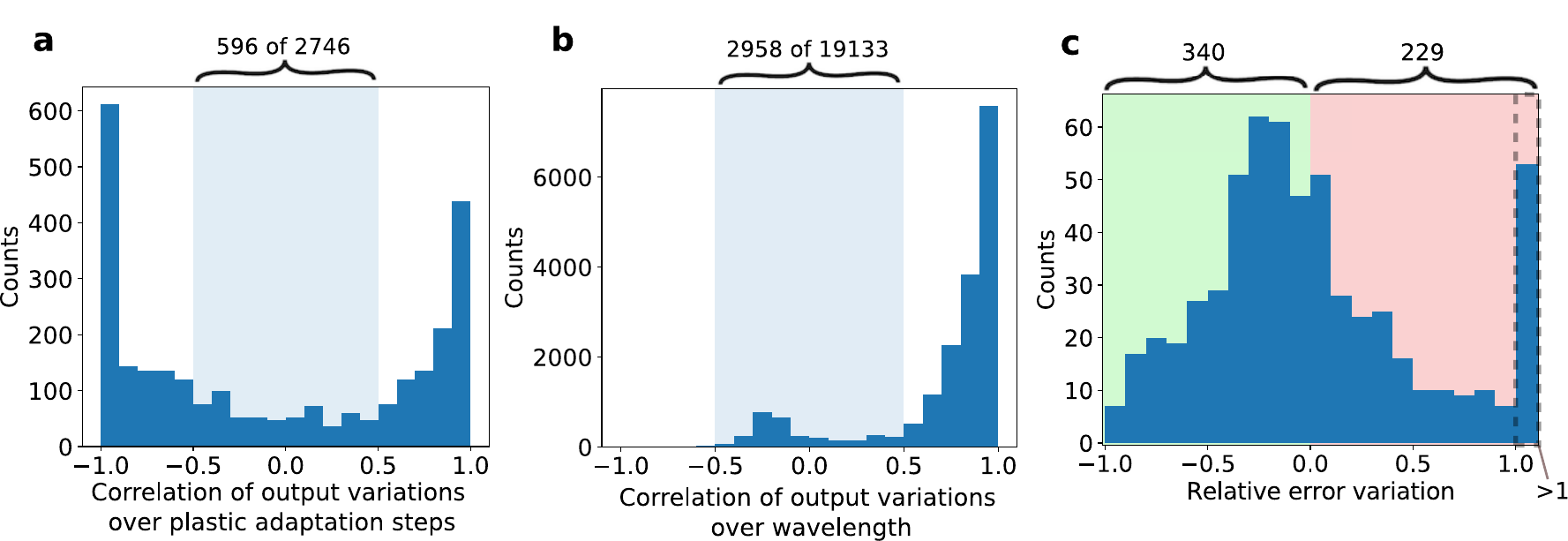}
  \centering
  \caption{\textbf{Results of plasticity investigation.} \textbf{a, b}
Histograms of the correlation coefficients between pairs of output feature variations, respectively for different PA steps and wavelengths. \textbf{c} Histogram of the error rate relative variations w.r.t. the initial error, in each sequence of PA steps.}
  \label{fig:historgrams}
\end{figure}

We now aim to give an overview of how differently the network output is modified by the different input waveforms (i.e. due to plastic adaptation (PA) steps from different classes). In particular,  by `network output' we mean here the output feature vector extracted from the signals at different physical output ports of the PPRRNN. Namely, as explained in the main text, for a given input waveform (i.e. a sample from the machine learning perspective), we consider the energy of the last pulse in the corresponding waveforms at each output port. Thus, the considered output features vector has as many scalar elements as the number of physical output ports.
To understand how the output features vector changes with different PA steps, we calculate the Pearson correlation coefficient between each pair of median variations of elements of the output feature vector,  over the PA step sequence. E.g., looking at Fig. 2e in the main text, the correlation is calculated between the two vectors represented in the first and the second box on the first row, then between the two vectors represented in the first and the third box in the same row, and so on until all the pairs are covered. This is repeated for each measured PA step sequence. Obviously, here we consider only the measurement sessions with multiple output ports (first three in Table 2 in the main text). Then, we represent all these correlation values in a histogram (Fig. \ref{fig:historgrams}a). We can notice that, indeed, a significant portion of the investigated PA step pairs provide output feature variations with low correlations. In particular, 596 correlation coefficients of a total of 2746 are between -0.5 and 0.5 (this correlation range is arbitrarily chosen, to provide some intuitive insight). These low correlations are desired and show that the PPRRNN plastic response exhibits \textit{richness}, as it can vary significantly depending on the input history.  On the other hand, it should be noticed that a large number of pairs are highly anti-correlated, even exceeding the number of highly correlated pairs. This means that the variations due to a PA step are often reverted by another PA step in the same series. In other words, this shows that the plastic weights configuration in the PPRRNN can be often reset by a single PA step. 

We then performed a similar analysis by calculating the output feature variation correlation for different wavelength pairs, in the same measurement session and for a fixed PA step. E.g., looking at Fig. 2e in the main text, the correlation between the two vectors represented in the two boxes along the first column is estimated, the along the second column, and so on. Also in this case, we can see that a significant portion of variation pairs present low correlation (2958 of a total of 19133 have correlation between -0.5 and 0.5). This means that PA can present very different outcomes for different wavelengths of the input optical signal. On the other hand, it is not surprising that most of the pairs are highly correlated, since the minimum wavelength difference is only $\SI{0.005}{\nano\meter}$ and the maximum is $\SI{0.1}{\nano\meter}$, to be compared with the width (at half minimum) of a MRR resonance dip, which is larger than $\SI{0.2}{\nano\meter}$.

We now discuss our analysis of the machine learning (ML) classification results obtained for each measurement session, each wavelength and each PA step. Mainly, we are interested in seeing how different achieved plastic weights configurations result in different ML performance, i.e. in different performance of the PPRRNN when employed to provide useful data representation to be fed to a linear classifier. In addition to the three measurement sessions considered before in this section (first three in Table 2 in the main text), our analysis now comprises also measurement sessions where a single output port and a single wavelength are used, so to allow more (30) PA steps without exceedingly increasing the measurement duration (in Table 2 in the main text, from session 4 to 12).
In order to provide an overall picture of the impact of the PA step in all the considered sequences, we show a histogram (Fig. \ref{fig:historgrams}c) of the error rate variations relative to the corresponding initial error rate value, due to each single PA step in all the measurement sessions. (This is different from Fig. 2g in the main text, which only considers the minimum over all PA steps, instead of every PA step individually.) It can be noticed that the value distribution is centered at negative relative error variations, showing that it was more frequent that a PA step resulted in a better ML performance w.r.t. the starting value, rather than the opposite. The counts in the negative relative error variation range are 340, against the 229 in the positive range. It should be considered that, since the plotted values are relative variations w.r.t. the initial value and they cannot be less than -1 (the error rate always being a positive number), the mean of the distribution is skewed towards positive values, which explains the highly populated tail consisting of values larger than 1. 

\bibliographystyle{unsrt}  
\bibliography{references}